\documentclass[preprint,showpacs,preprintnumbers,amsmath,amssymb]{revtex4-1}
\usepackage{graphics,epsfig,subfigure}
\usepackage{diagbox}
\usepackage[usenames]{color}
\usepackage[colorlinks,
linkcolor=blue,
anchorcolor=red,
citecolor=red
]{hyperref}
\usepackage{color}
\usepackage{graphicx}
\usepackage{amsmath}
\usepackage{amsfonts}
\usepackage{amssymb}
\usepackage{txfonts}
\usepackage{indentfirst}
\usepackage{booktabs}

\begin{document}
	\renewcommand{\baselinestretch}{1.3}
	\newcommand\beq{\begin{equation}}
		\newcommand\eeq{\end{equation}}
	\newcommand\beqn{\begin{eqnarray}}
		\newcommand\eeqn{\end{eqnarray}}
	\newcommand\nn{\nonumber}
	\newcommand\fc{\frac}
	\newcommand\lt{\left}
	\newcommand\rt{\right}
	\newcommand\pt{\partial}

\title{\Large{\bf Hayward spacetime with axion scalar field}}

\author{ Jun-Ru Chen and  Yong-Qiang Wang\footnote{yqwang@lzu.edu.cn, corresponding author}
	}
	
\affiliation{$^{1}$ Lanzhou Center for Theoretical Physics, Key Laboratory of Theoretical Physics of Gansu Province, School of Physical Science and Technology, Lanzhou University, Lanzhou 730000, China\\
 $^{2}$ Institute of Theoretical Physics $\&$ Research Center of Gravitation, Lanzhou University, Lanzhou 730000, China}
	\date{\today}

\begin{abstract}
In this work, we investigate a static spherically symmetric system in which Einstein gravity is minimally coupled with a self-interacting complex scalar field and a nonlinear electromagnetic field, referred to as Hayward axion stars. Employing numerical methods, we find that it essentially describes axion stars with the magnetic charge. In the absence of magnetic charge and with only the scalar field present, the system reduces to axion stars. We discover that when the magnetic charge $q$ exceeds a critical value, extreme solutions with frequencies $\omega$ approaching zero can be found and the critical horizon emerges. Within this horizon, the scalar field and energy density are highly concentrated and decrease precipitously at its boundary.
The time component of the metric function approaches zero within this region, indicating that gravity is extremely intense, and time nearly ceases to flow. To an observer at infinity, the star appears to be frozen, hence we refer to these extreme solutions exhibiting a critical horizon as Hayward axion frozen stars. Furthermore, it is important to note that as $\omega \rightarrow 0$, the mass of the Hayward axion frozen star becomes independent of the decay constant and is only determined by the magnetic charge. Additionally, we find that the frozen star solutions possess two light rings. With an increase in the magnetic charge, these light rings move outward, while changes in the decay constant have little effect on their positions.
\end{abstract}	

\maketitle
 \newpage
\section{Introduction} 
General relativity, introduced by Einstein in 1916~\cite{Einstein:1916vd}, predicts the existence of black holes. These are regions in spacetime where the gravitational field is so intense that even light cannot escape. The mechanism of black hole formation was elucidated by Oppenheimer and Snyder, which occurs through the gravitational collapse of a massive star when its internal pressure is no longer able to withstand its own gravity. This process continues indefinitely if the mass of the star cannot be reduced below that of the sun through mechanisms such as rotation and thermal radiation, ultimately leading to the formation of a black hole~\cite{Oppenheimer:1939ue}. Observations of binary black hole mergers producing gravitational waves by LIGO~\cite{LIGOScientific:2016sjg,
LIGOScientific:2016kms,
LIGOScientific:2016aoc} and images of the shadow of a supermassive black hole by the Event Horizon Telescope~\cite{EventHorizonTelescope:2019dse,
EventHorizonTelescope:2019uob,
EventHorizonTelescope:2019jan,
EventHorizonTelescope:2019ths,
EventHorizonTelescope:2019pgp,
EventHorizonTelescope:2019ggy,
EventHorizonTelescope:2022wkp,
EventHorizonTelescope:2022apq,
EventHorizonTelescope:2022wok,
EventHorizonTelescope:2022exc,
EventHorizonTelescope:2022urf,
EventHorizonTelescope:2022xqj} have further confirmed the existence of black holes. Theoretical studies of black holes have consistently attracted significant interest since 1916 when Schwarzschild discovered a spherically symmetric vacuum solution for a static, uncharged black hole by solving the field equations of general relativity~\cite{Schwarzschild:1916uq}. This solution was later extended to include electric charge by Reissner and Nordstrom~\cite{Reissner:1916cle, nordstrom1918mass}. Further theoretical research of black holes was conducted by Kerr, who discovered the solution describing the gravitational field of rotating black holes~\cite{Kerr:1963ud}. Subsequently, Newman et al. extended the Kerr solution to include charge~\cite{Newman:1965my}.

However, a fundamental challenge with these black hole solutions is the presence of spacetime singularities within them, where physical quantities become infinite, leading to the breakdown of physical laws in these regions. The singularity theorems proposed by Penrose and Hawking indicate that under general relativity, spacetime singularities arise in the solutions of Einstein's equations when the strong energy condition and globally hyperbolic spacetime assumptions are considered~\cite{Penrose:1964wq, Hawking:1966sx,
Hawking:1966jv,
Hawking:1967ju}. Furthermore, the weak cosmic censorship hypothesis posits that such singularities are enveloped by event horizons~\cite{Wald:1997wa, Jhingan:1999vj}. 
The event horizon serves as a boundary that separates the black hole's interior from the exterior, thereby severing any causal connections between the two.
Consequently, any pathologies within the singular region do not influence the exterior, ensuring that the external descriptions of general relativity remain self-consistent. However, singularities are essentially mathematical constructs, that do not manifest in the real physical world, given that matter cannot be infinitely compressed into a point.

In order to solve the problem of the singularity inside black holes,  Bardeen developed a solution now known as the Bardeen black hole~\cite{bardeen1968non}. This solution assumes a suitable mass function that employs nonlinear magnetic monopoles to regularize the central singularity~\cite{Ayon-Beato:2000mjt}. These solutions are regular everywhere and represent the first regular black hole models with event horizons that comply with the weak energy condition. Following this development, Hayward also provided a metric for a class of regular black holes~\cite{Hayward:2005gi}. The exploration of regular black holes lead to the development of a variety of solutions that further extend our understanding of regular black holes~\cite{Dymnikova:1992ux,
Mars:1996khm,
Ayon-Beato:1998hmi,
Ayon-Beato:1999kuh,
Bronnikov:2000yz,
Bronnikov:2000vy,
Dymnikova:2003vt,
Dymnikova:2004zc,
Ayon-Beato:2004ywd,
Uchikata:2012zs,
Balart:2014cga,
Balart:2014jia,
Frolov:2016pav,
PoncedeLeon:2017usu,
Sajadi:2017glu}.

Additionally, boson stars have garnered significant interest as models without singularities. In the mid-1950s, Wheeler constructed a system by coupling gravity with classical electromagnetic fields in general relativity, which he termed geons~\cite{Wheeler:1955zz, Power:1957zz}.
Geons are localized energy configurations without singularities formed by the interaction of electromagnetic waves with their own gravitational field.
 However, this configuration is unstable and tends to undergo gravitational collapse when perturbed.
Subsequently, Kaup and Ruffini et al. replaced the electromagnetic field with a scalar field, constructing stable solutions~\cite{Kaup:1968zz, Ruffini:1969qy}. These soliton configurations formed by their own gravitational binding are known as boson stars.

Recent study has investigated  the complex scalar field coupled with Bardeen spacetime to construct the Bardeen boson star model~\cite{Wang:2023tdz}. 
This model describes Bardeen spacetime in the absence of a matter field and a boson star when there is no magnetic charge. 
Notably, in the presence of the matter field, Bardeen boson stars do not have the event horizon but only a critical horizon. 
When the magnetic charge exceeds a critical value, solutions with frequencies approaching zero appear, leading to a critical radius $r_c$ where the scalar field is mainly distributed. 
At this critical radius, the metric component $1/g_{rr}$ approaches zero without actually reaching zero, hence the surface at $r=r_c$ is referred to as the critical horizon.
Within the critical horizon, the metric component $g_{tt}$ is close to zero, and the flow of time is significantly slowed. 
From a distance, matter falling into such stars appears to be frozen at the critical horizon, thus these stars are also known as frozen stars.
The early research on frozen stars was provided by Zel'dovich and Novikov~\cite{Zeldovich and Novikov}, reflecting the phenomenon that light escaping from the region near the gravitational radius to a distant observer takes an infinitely long time, making the star appear frozen near its gravitational radius~\cite{Ruffini:1971bza}. Inspired by this idea, solutions for frozen stars without event horizons were also found in~\cite{Huang:2023fnt,Yue:2023sep,Ma:2024olw}.

The purpose of this paper is to investigate axion fields within Hayward spacetime. By coupling the Einstein-Klein-Gordon theory with the nonlinear electrodynamics proposed by S. Hayward, we construct Hayward axion stars. 
The axion is initially introduced by Peccei and Quinn in the 1970s~\cite{Peccei:1977hh} to solve the strong CP problem in particle physics~\cite{Jackiw:1976pf,
Callan:1979bg}, and it is considered a viable candidate for dark matter~\cite{Davidson:2016uok,
Klaer:2017ond}. 
Our analysis involves obtaining numerical solutions for Hayward axion stars under various decay parameters and analyzing their physical properties.
 We find that these models, which are essentially axion stars with magnetic charge, do not possess event horizons. Under specific conditions, frozen star solutions will appear.
Additionally, we explore the presence of light rings around Hayward axion stars and analyze the influence of magnetic charge and decay constant on the position of light rings under the frozen star solutions.

The organization of this paper is as follows. In Section~\ref{sec2}, we introduce the Hayward axion star which minimally couples gravity with electromagnetic and the matter field. In Section~\ref{sec3}, we give the numerical method and discuss the boundary conditions. In Section~\ref{sec4}, we provide the numerical results for Hayward axion stars and analyze their physical properties. Finally, the conclusion and discussion of the paper are presented in the last section.

       \section{The model} \label{sec2}
 We consider the minimal coupling of Einstein gravity with a self-interacting complex scalar field and a nonlinear electromagnetic field. The action for this model is expressed as
         \begin{equation}\label{action}
  S=\int\sqrt{-g}d^4x\left(\frac{R}{16 \pi G}+\mathcal{L}^{(A)}+\mathcal{L}^{(H)}\right) \ ,
\end{equation}
where $\mathcal{L}^{(A)}$ and $\mathcal{L}^{(H)}$  are the Lagrangian densities for the axion field~\cite{Liebling:2012fv} and the electromagnetic field~\cite{Fan:2016hvf}, respectively, as given by
    \begin{eqnarray}
    \mathcal{L}^{(A)} &= & - g^{\alpha\beta} \partial_{\alpha} \Psi^* \partial_{\beta} \Psi - V(|\Psi|^2) 
 \ ,\\
 \mathcal{L}^{(H)} &= &- \frac{ 3}{ 2 s } \frac{ (2 q^2 {\cal F})^{3/2}}{\left(  1 + ( 2 q^2 {\cal F})^{3/4}\right)^2}
\ ,
\end{eqnarray}
here $R$ represents the curvature scalar, $G$ is the Newtonian gravitational constant, and ${\cal F}$ denotes the Maxwell scalar defined as ${\cal F} = \frac{1}{4}F_{ab} F^{ab}$, with $F_{ab} = \partial_{a} A_{ b} - \partial_{b} A_{ a}$ representing the electromagnetic field tensor associated with the vector potential  $A_{ a}$. The complex scalar field is denoted by $\Psi$, and $V(|\Psi|^2)$ is the axion potential. 
The variables $q$ and $s$ are two independent positive parameters, with $q$ representing the magnetic charge and $s=q^3/2M$. Upon varying Eq.~\eqref{action} with respect to the metric, electromagnetic field, and scalar field, we derive the following equations of motion
\begin{eqnarray} \label{eq:motion}
R_{\alpha\beta}-\frac{1}{2}g_{\alpha\beta}R- 8\pi G (T^{(1)}_{\alpha\beta}+T^{(2)}_{\alpha\beta})&=&0 \ ,  \nonumber\\
\bigtriangledown_{\alpha} \left(\frac{ \partial {\cal L}^{(1)}}{ \partial {\cal F}}  F^{\alpha\beta}\right) &=& 0\ ,    \\
\Box \Psi -\dfrac{\partial V}{\partial |\Psi|^2} \Psi &=& 0\ , \nonumber
\end{eqnarray}
with
\begin{equation}
T^{(1)}_{\alpha\beta} =- \frac{ \partial {\cal L}^{(1)}}{ \partial {\cal F}} F_{\alpha \gamma} F_{ \beta }^{\;\;\gamma} + g_{\alpha\beta} {\cal L}^{(1)}\ ,
\end{equation}
\begin{equation}
T^{(2)}_{\alpha\beta} = \partial_\alpha \Psi^*\partial_\beta \Psi + \partial_\beta \Psi^*\partial_\alpha \Psi - g_{\alpha\beta}\left[\frac{1}{2}g^{\alpha\beta}\left(\partial_\alpha \Psi^*\partial_\beta \Psi + \partial_\beta \Psi^*\partial_\alpha \Psi\right) + V\right]\ .
\end{equation}
The action Eq. \eqref{action} exhibits two special cases. Firstly, in the case where $q=0$ but the complex scalar field is present, it delineates the Einstein-Klein-Gordon theory, and the solutions in this case degenerate into spherically symmetric axion stars~\cite{Guerra:2019srj}. Secondly, when the complex scalar field vanishes but the magnetic charge persists, the model degenerates into the Hayward model with the minimal coupling between general relativity and nonlinear electrodynamics~\cite{Hayward:2005gi}. In the static spherically symmetric model of Hayward spacetime, the line element of the Hayward model is given by
\begin{equation}
ds^2 = -f(r) dt^2 + f(r)^{-1} dr^2 + r^2 d\Omega\ ,
\end{equation}
where $d\Omega =d\theta^2 + \sin^2(\theta) d\varphi^2$ and the Hayward metric function is
\begin{equation}
f(r) = 1 - \frac{1}{s} \frac{q^3 r^2}{ ( r^3 + q^3 ) }\ .
\end{equation}
The metric function $f(r)$ has a local minimum at $r = 2^{1/3} q$. The equation $f(r)=0$ yields a real root when $s=s_c=\frac{2^{2/3} q^2}{3}$, which corresponds to the extreme case of Hayward black holes with horizon degeneracy. When $s>s_c$, no solutions exist, indicating the absence of a horizon. Conversely, when $s<s_c$, two real roots are present, signifying the existence of two horizons.

According to Noether's theorem, the invariance of the action for a complex scalar field under U(1) transformations, represented as $\Psi \to \Psi e^{i\alpha}$ with $\alpha$ held constant, implies the existence of a conserved current
\begin{equation}
	j^{\alpha} = -i \left( \Psi^* \partial^{\alpha} \Psi - \Psi \partial^{\alpha} \Psi^* \right) \ . 
	\label{Currents}
\end{equation}
By integrating the timelike component of the conserved current over the spacelike hypersurface $\Omega$, one obtains a conserved quantity, the Noether charge associated with the complex scalar field
\begin{equation}
	Q =  \int_\Omega j^t \ .
	\label{Charge}
\end{equation}
In asymptotically flat spacetimes, the ADM mass is determined by the integral of the energy density $\rho$, where
\begin{equation}
	M=\int  dr r^2 \rho \ .	
\end{equation}

We are interested in spherically symmetric static configurations, which can be described using the following metric ansatz
\begin{equation}
	ds^2 = -e^{u(r)} dt^2 + e^{v(r)} dr^2 + r^2 \left(d \theta^2+\sin^2 \theta d \varphi^2 \right)\ ,
	\label{metric}
\end{equation}
where the functions $u(r)$ and $v(r)$ are dependent on the radial variable $r$. Using the ansatzes for both the electromagnetic field and the scalar field
\begin{equation}\label{field}
   A_\varphi= q \cos(\theta),\;\;\; \Psi=\psi(r) e^{-i \omega t}\ ,
\end{equation}
where $\psi$ is a real function dependent on $r$ and $\omega > 0$ denotes the frequency of the scalar field. The non-zero components of the electromagnetic field tensor are $F_{34}=-F_{43}=-q \sin(\theta)$, thus the Maxwell scalar is ${\cal F}=q^2/{2r^4}$.

The axion potential is given by~\cite{Guerra:2019srj,
GrillidiCortona:2015jxo}
\begin{equation}
	V(\psi) = \frac{2 \mu^2 f_a^2}{B} \left[ 1 - \sqrt{1 - 4 B \sin^2 \left( \frac{\psi}{2 f_a} \right)} \right] \ ,
	\label{Potential}
\end{equation}
where $B=\frac{m_u m_d}{\left(m_u+m_d\right)^2}\approx0.22$  is associated with the ratio of up and down quark masses $m_u/m_d \approx 0.48$, $\mu$ represents the mass of the axion, and $f_a$ is the decay constant of the axion field. Expanding the axion potential at $\psi=0$ yields
\begin{equation}
	V(\psi) = \mu^2 \psi^2 - \left( \frac{3B-1}{12} \right) \frac{ \mu^2}{f_a^2} \psi^4 + \dots \ .
	\label{UnfoldingPotential}
\end{equation}
Thus, it can be observed that when $f_a \gg \psi$, the self-interaction term in the axion potential approaches zero. Under these conditions, the axion star model decays into the mini-boson star~\cite{Kaup:1968zz, Ruffini:1969qy}.

By applying the axion potential Eq.\eqref{Potential} and substituting the metric ansatz given in Eq.\eqref{metric} and the ansatz for the scalar and electromagnetic fields given in Eq.\eqref{field} into the motion equation Eq.~\eqref{eq:motion}, we derive the following equations for $u$, $v$, and $\psi$
\begin{equation}
\begin{aligned}
 u^{\prime}  = & \frac{1-e^{u}}{r} +8 G \pi r\left(e^{u-v} \omega^2 \psi^2+\psi'^2\right)+
\\
&4 e^{u} G \pi r\left(\frac{3 q^6}{\left(q^3+r^3\right)^2 s}-\frac{4 f_a^2 \mu^2\left(-1+\sqrt{1-2 B +2 B \operatorname{\cos}\left(\frac{\psi}{f_a}\right)}\right)}{B }\right)
\ , 
\end{aligned}
\end{equation}
\begin{equation}
\begin{aligned}
 v^{\prime} = & \frac{-1+e^{u}}{r}+8 G \pi r\left(e^{u-v} \omega^2 \psi^2+\psi'^2\right) +
 \\
&4 e^{u} G  \pi r\left(-\frac{3 q^6}{\left(q^3+r^3\right)^2 s}+\frac{4 f_a^2 \mu^2\left(-1+\sqrt{1-2 B +2 B  \operatorname{\cos}\left(\frac{\psi}{f_a}\right)}\right)}{B }\right)
\ , 
\end{aligned}
\end{equation}
\begin{equation}
\begin{aligned}
\psi^{\prime \prime} =  \frac{e^u f_a \mu^2 \operatorname{\sin}\left(\frac{\psi}{f_a}\right)}{\sqrt{1-2 B +2 B  \operatorname{\cos}\left(\frac{\psi}{f a}\right)}}-e^{u-v} \omega^2 \psi-\frac{1}{2}\left(\frac{4}{r}-u^{\prime}+v^{\prime}\right) \psi^{\prime}
\ . 
\end{aligned}
\end{equation}
Here, prime denotes differentiation concerning the radial coordinate $r$. Through the imposition of suitable boundary conditions, these three ordinary differential equations can be effectively solved.

       \section{Numerical method} \label{sec3}
In order to numerically solve the equations in Sec.~\ref{sec2}, it is essential to specify appropriate boundary conditions. At the origin, we impose regularity boundary for the metric and scalar field
\begin{equation}\label{Origin}
		u(0)=0 \ ,\qquad v(0)^{\prime}=0 \ , \qquad \psi^{\prime}(0)=0\ .
\end{equation}
And for asymptotically flat solutions, the asymptotic behavior of the metric and scalar field at infinity are required to be
\begin{equation}\label{infinity}
  \lim_{r\to \infty} v(r) = 0 \ ,\qquad  \lim_{r\to \infty} \psi(r) = 0 \ .
\end{equation}
To perform numerical integration, we introduce dimensionless quantities by re-scaling the variables as follows
\begin{equation}
r \rightarrow \mu r\ , \quad  M \rightarrow  M \mu ,\quad \omega \rightarrow \frac{\omega}{\mu}, \quad \psi \rightarrow \sqrt{4 \pi G} \psi
\end{equation}
Additionally, we use the following coordinate transformation
\begin{equation}
	x=\frac{r}{1+r} \ ,
	\label{}
\end{equation}
to map the range of the radial coordinate from  $r\in[0,\infty)$ to $x\in[0,1]$.
We employ the finite element method to numerically solve the system of ordinary differential equations. 
The number of grid points within the integration region 
$x \in[0,1]$ is 10000. The iterative method used is the Newton-Raphson method. To ensure the accuracy of the solutions, we maintain the relative error below $10^{-5}$.
In subsequent calculations, to maintain generality, we set $s=0.2$ and $G=1/4\pi$, while the decay constant 
$f_a$, magnetic charge $q$, and frequency $\omega$ are variable parameters.

       \section{Numerical Results} \label{sec4}
\begin{figure}[htbp]
    \centering
    \includegraphics[height=.28\textheight]{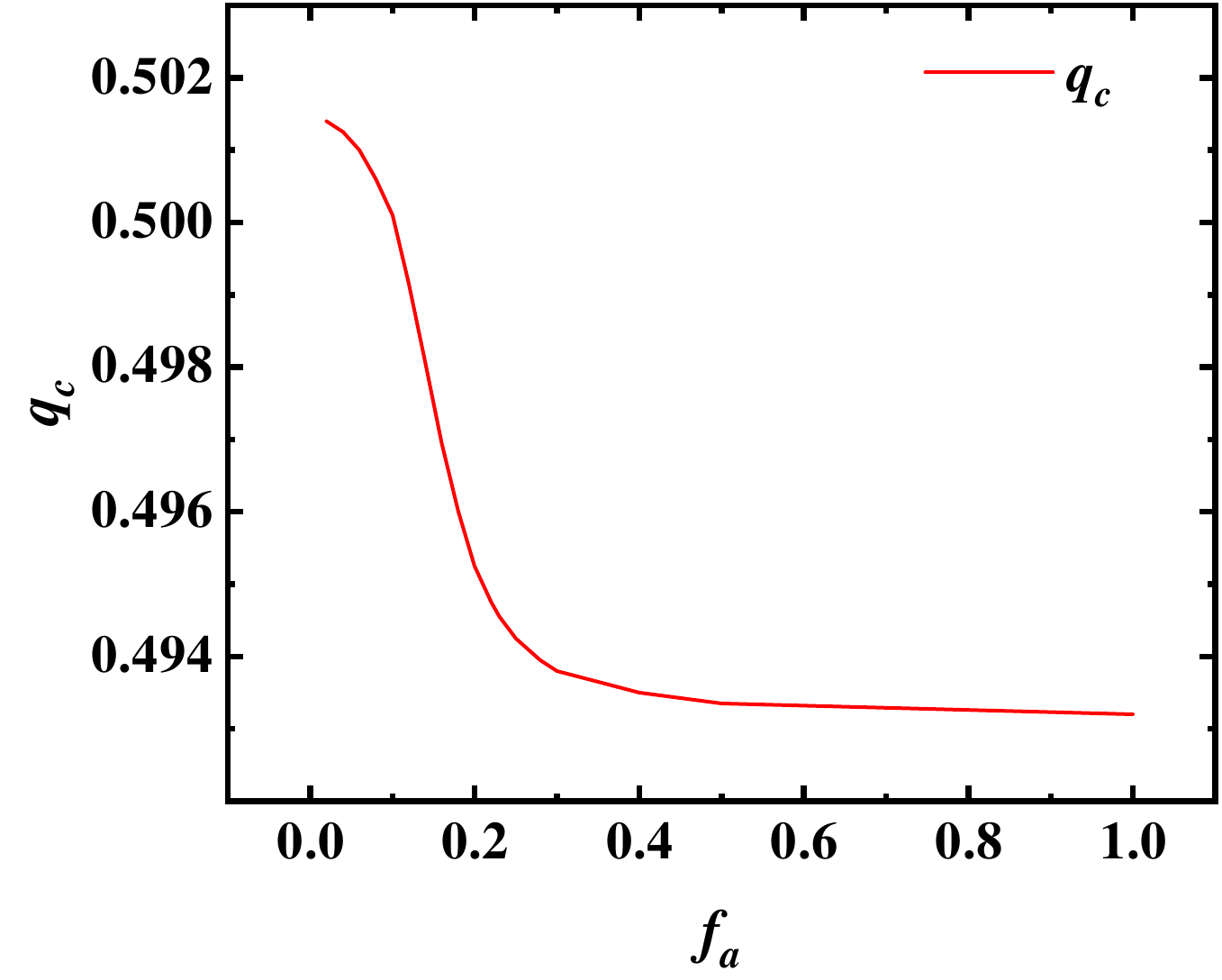}
    \caption{The critical magnetic charge $q_c$ as a function of the decay constant $f_a$.}
    \label{f_aq}
\end{figure}

       In this section, we explore the spherically symmetric axion star model influenced by a nonlinear electromagnetic field. 
        By solving the equations of motion, we only found solutions without event horizons.
        These solutions are essentially axion stars with a magnetic charge, which we refer to as Hayward axion stars.
        Our numerical solutions differentiate between two scenarios based on the magnitude of the magnetic charge. For low magnetic charge, where the magnetic charge $q$ is less than the critical magnetic charge $q_c$ ($q < q_c$), solutions with frequency approaching zero are not found. For high magnetic charge, where the magnetic charge $q$ exceeds the critical magnetic charge $q_c$ ($q \geq q_c$), we find extreme solutions with frequency $\omega \to 0$. 
         As shown in Fig.~\ref{f_aq}, the critical magnetic charge $q_c$ decreases as $f_a$ increases. This indicates that after the self-interaction is enhanced, the system needs a larger magnetic charge to reach the extreme solutions.  Moreover, the critical magnetic charge $q_c$ tends to stabilize for larger decay constants $f_a$, as the self-interaction of the axion field weakens and the potential function closely approximates a free potential.

 \subsection{Low magnetic charge ($q < q_c$)} 
 \begin{figure}[!t]
       	\centering
       	\includegraphics[width=0.45\textwidth]{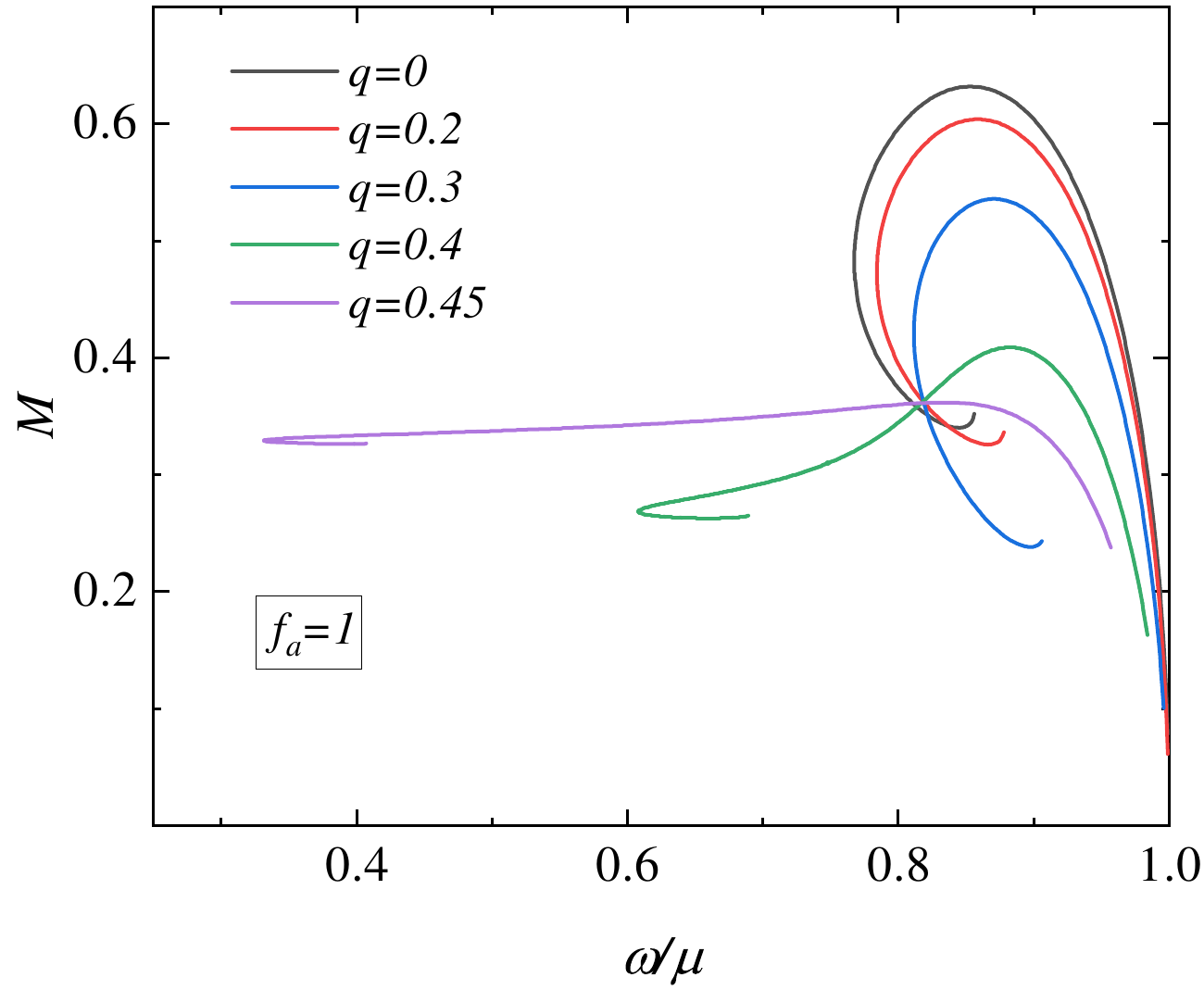}
       	\includegraphics[width=0.45\textwidth]{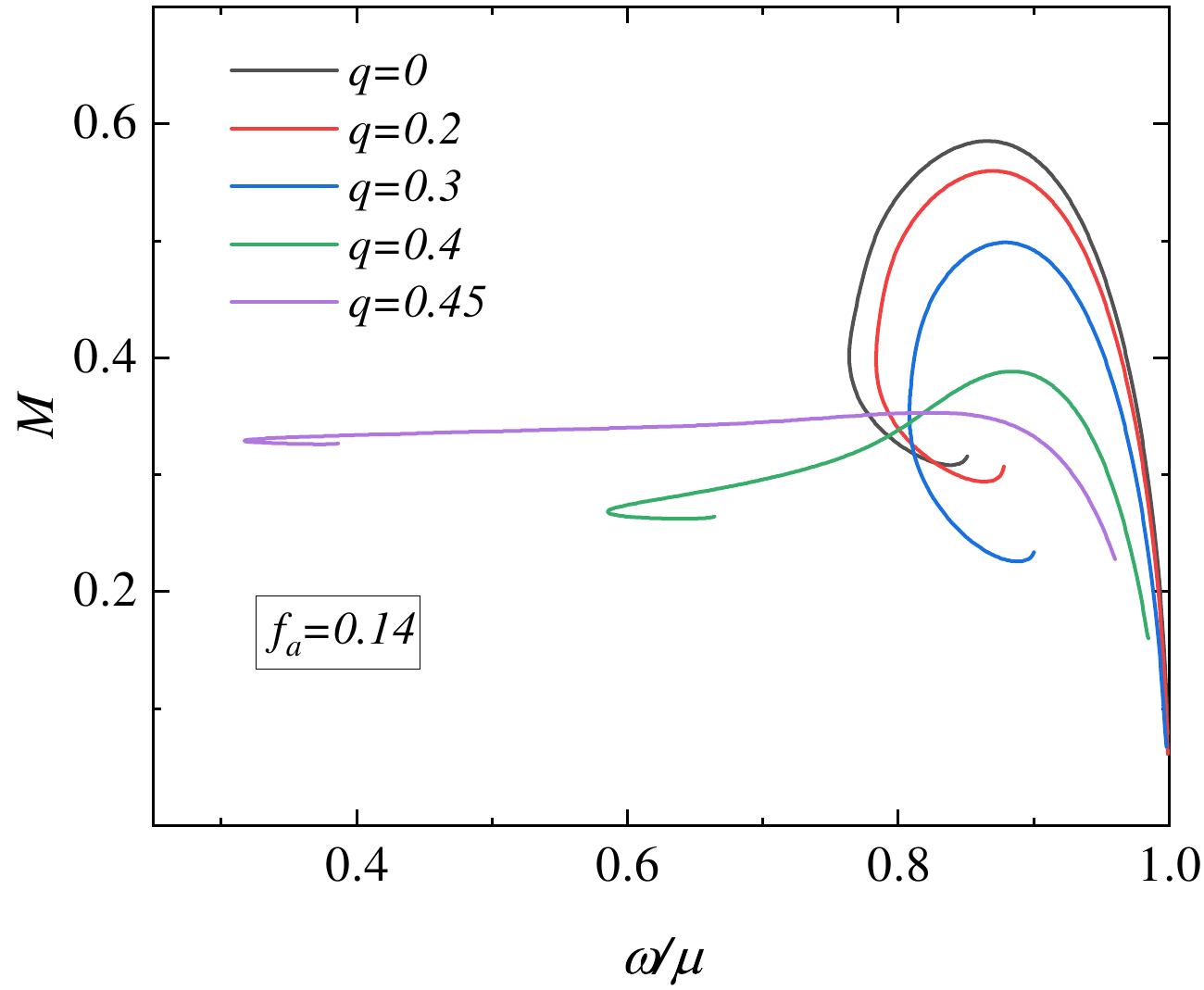}
       		\includegraphics[width=0.45\textwidth]{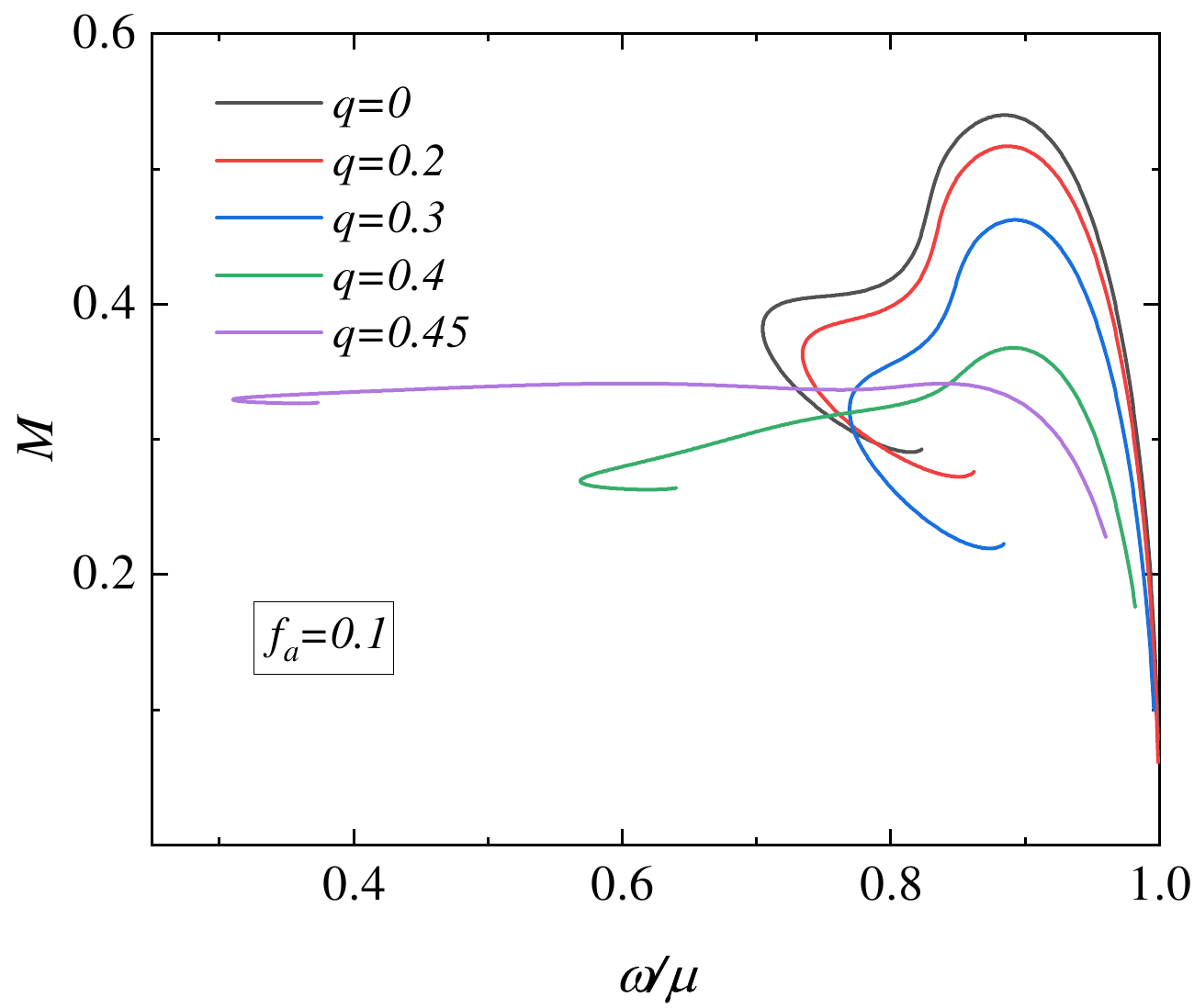}
       			\includegraphics[width=0.45\textwidth]{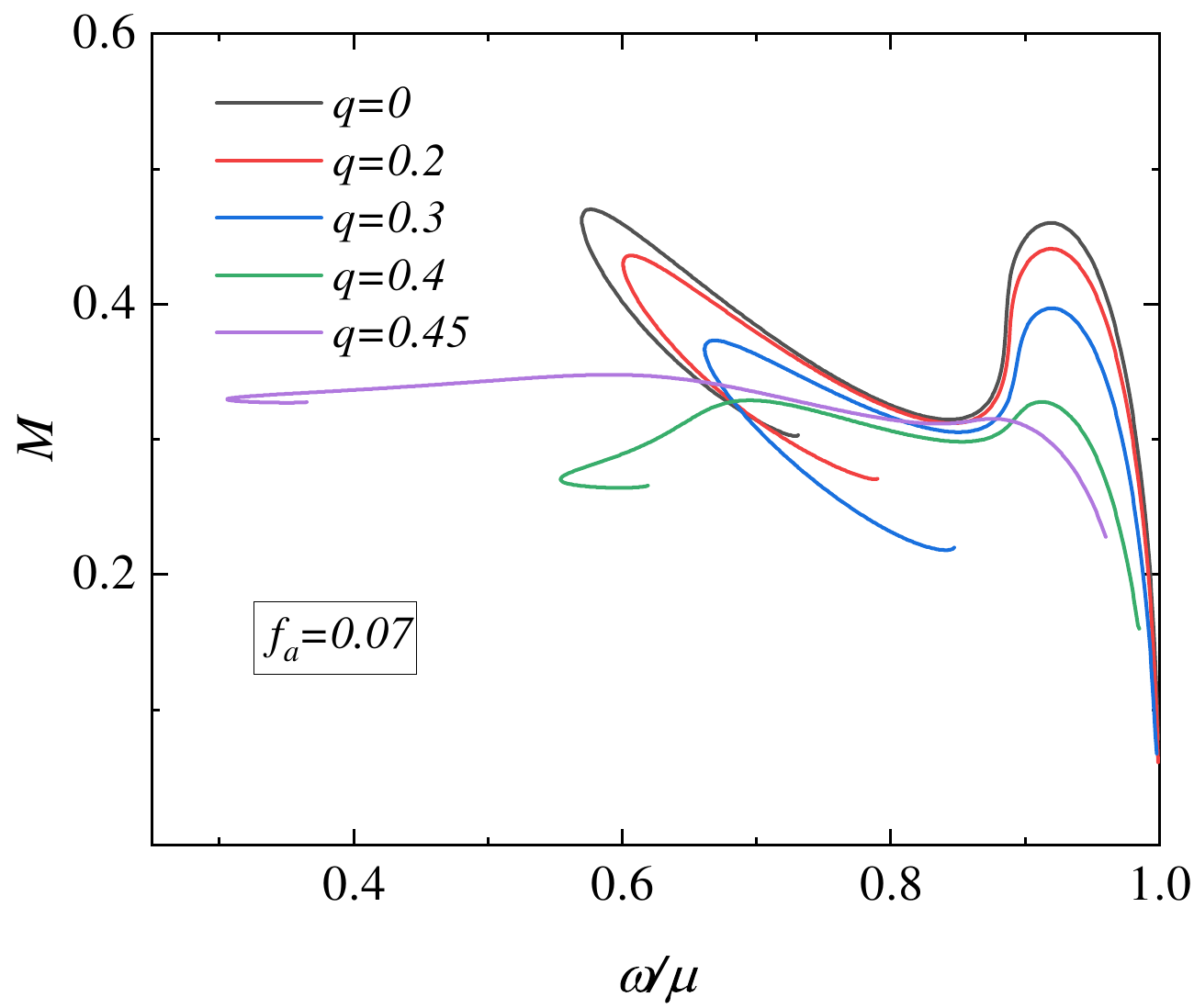}
       				\includegraphics[width=0.45\textwidth]{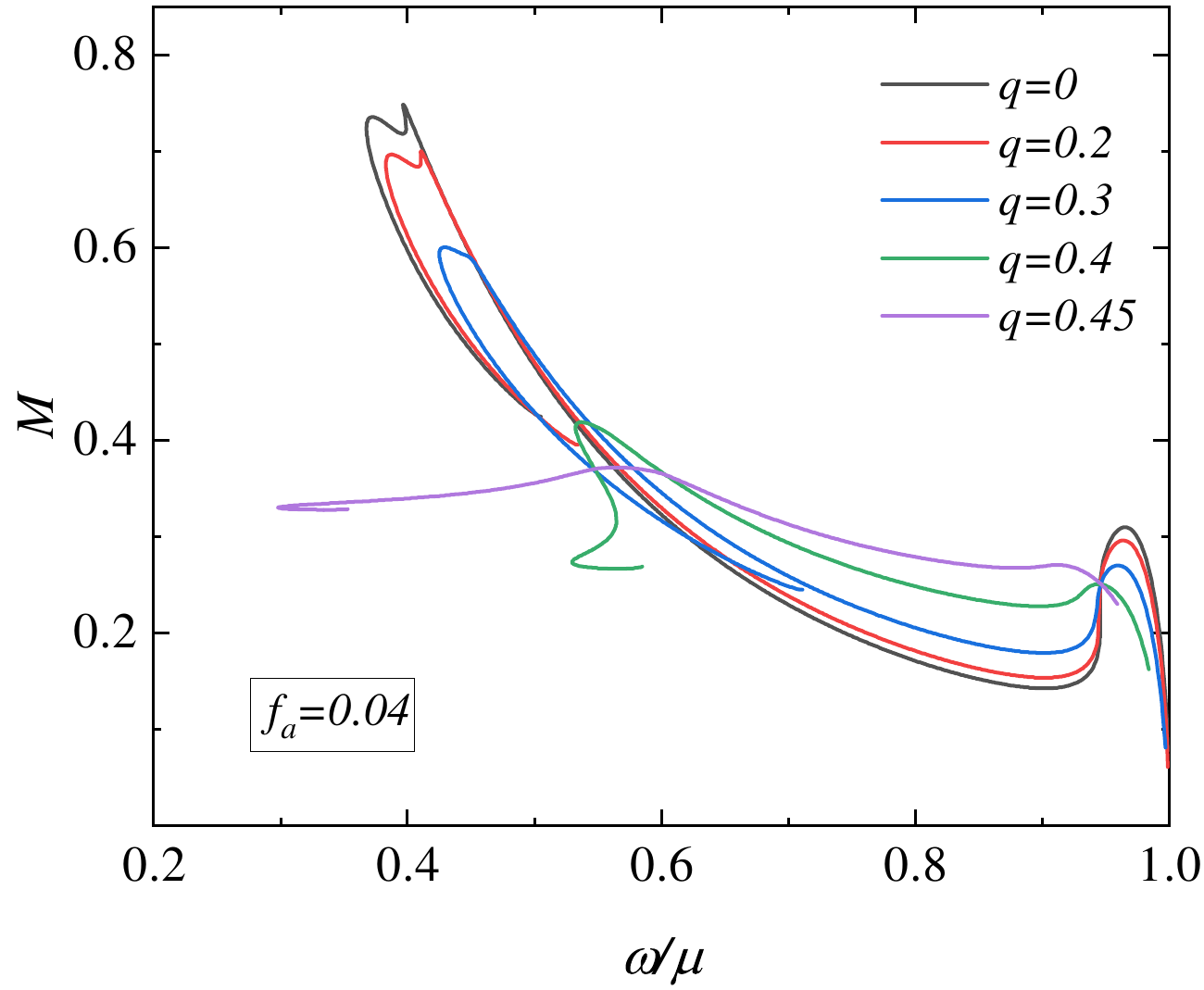}
       					\includegraphics[width=0.45\textwidth]{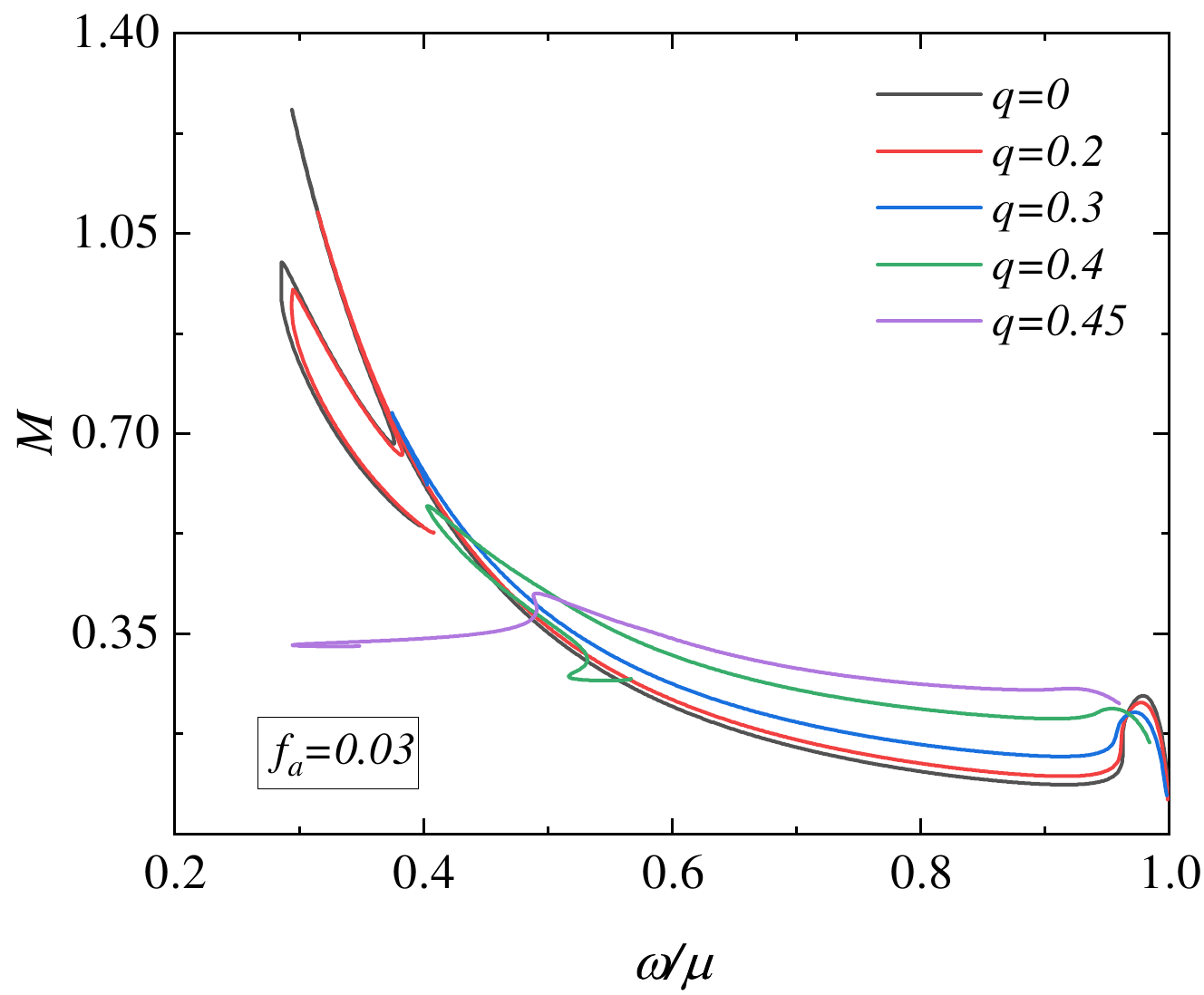}
       	\caption{The ADM mass $M$ as a function of frequency $\omega$ for hayward axion stars with low magnetic charge $q<q_c$ at decay constant $f_a=\{1,0.14,0.1,0.07,0.04,0.03\}$. Solid lines of the same color represent the same magnetic charge $q$, while the black solid line represents the solution for spherically symmetric axion stars without magnetic charge.}
       	\label{lowqmw}
       \end{figure}   
 \begin{figure}[!t]
   	\centering
   	\includegraphics[width=0.45\textwidth]{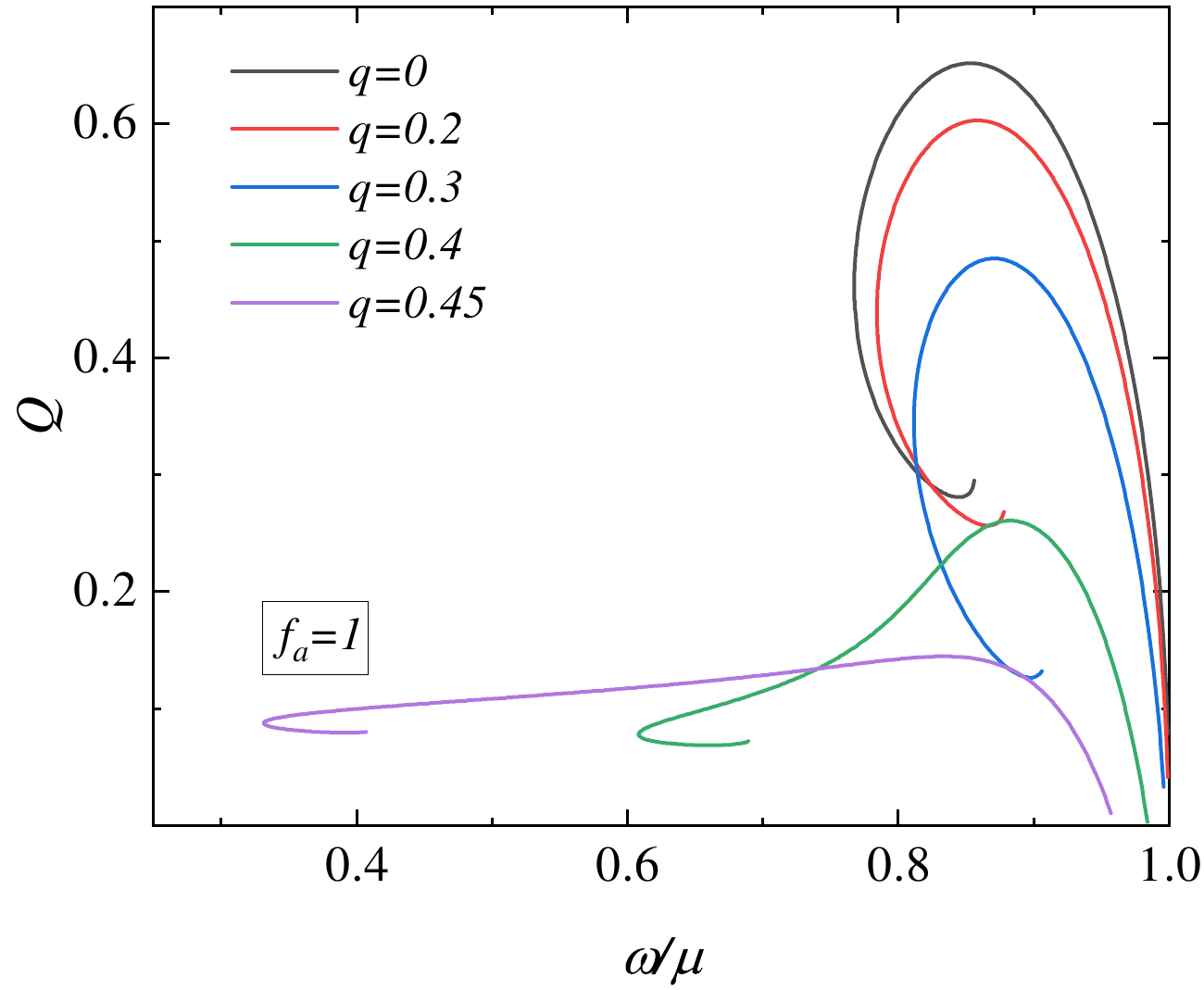}
   \includegraphics[width=0.45\textwidth]{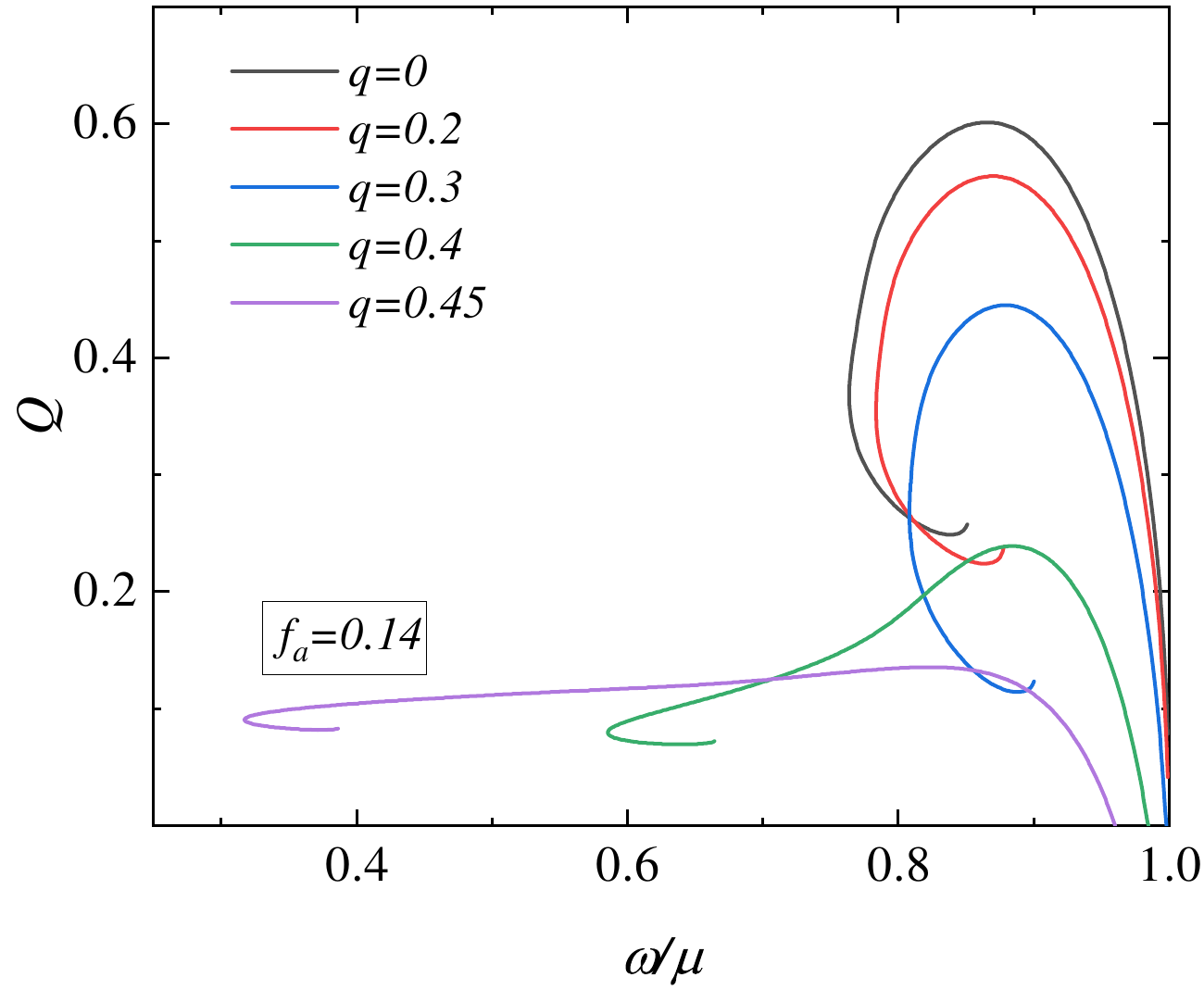}
   \includegraphics[width=0.45\textwidth]{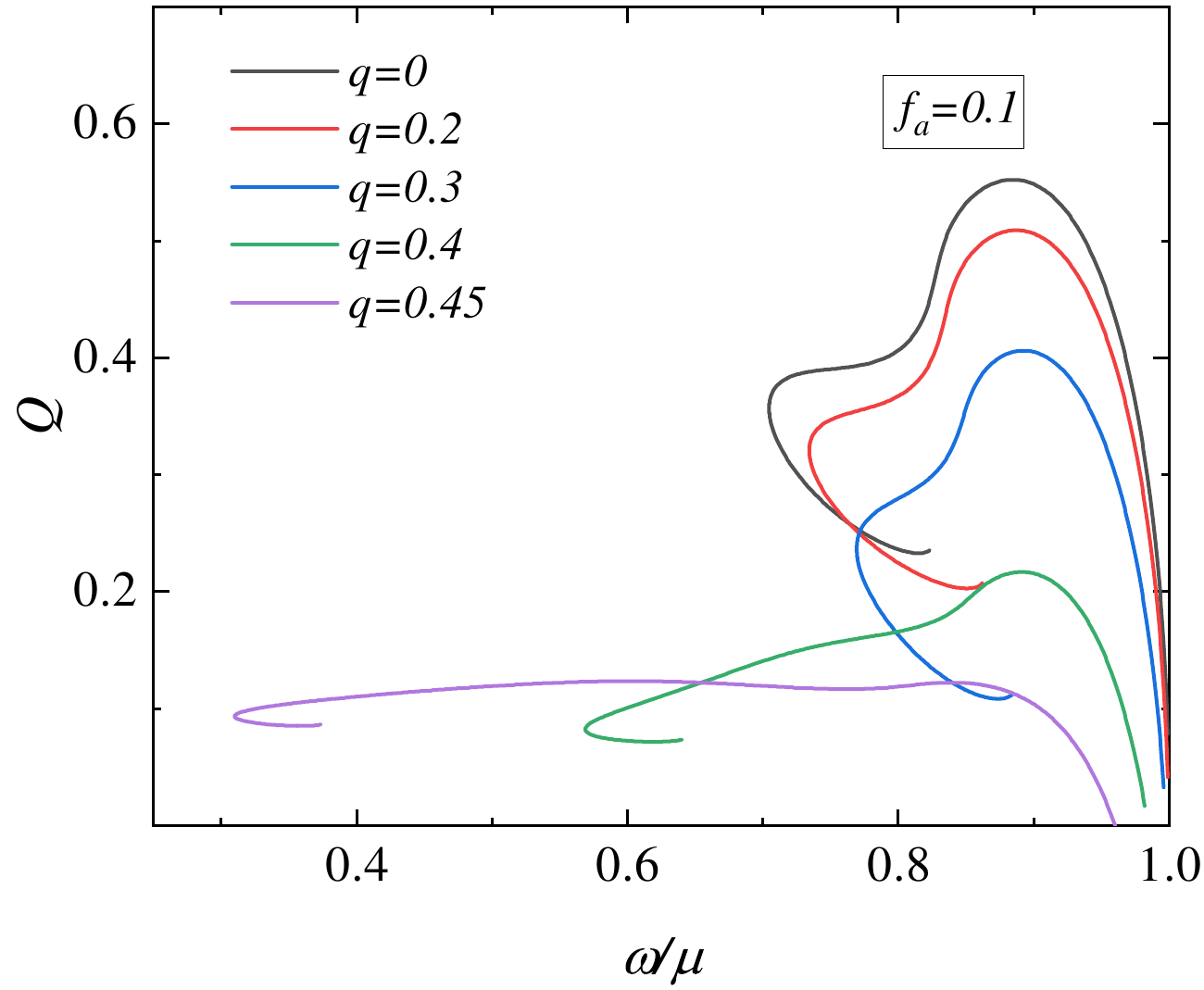}
   \includegraphics[width=0.45\textwidth]{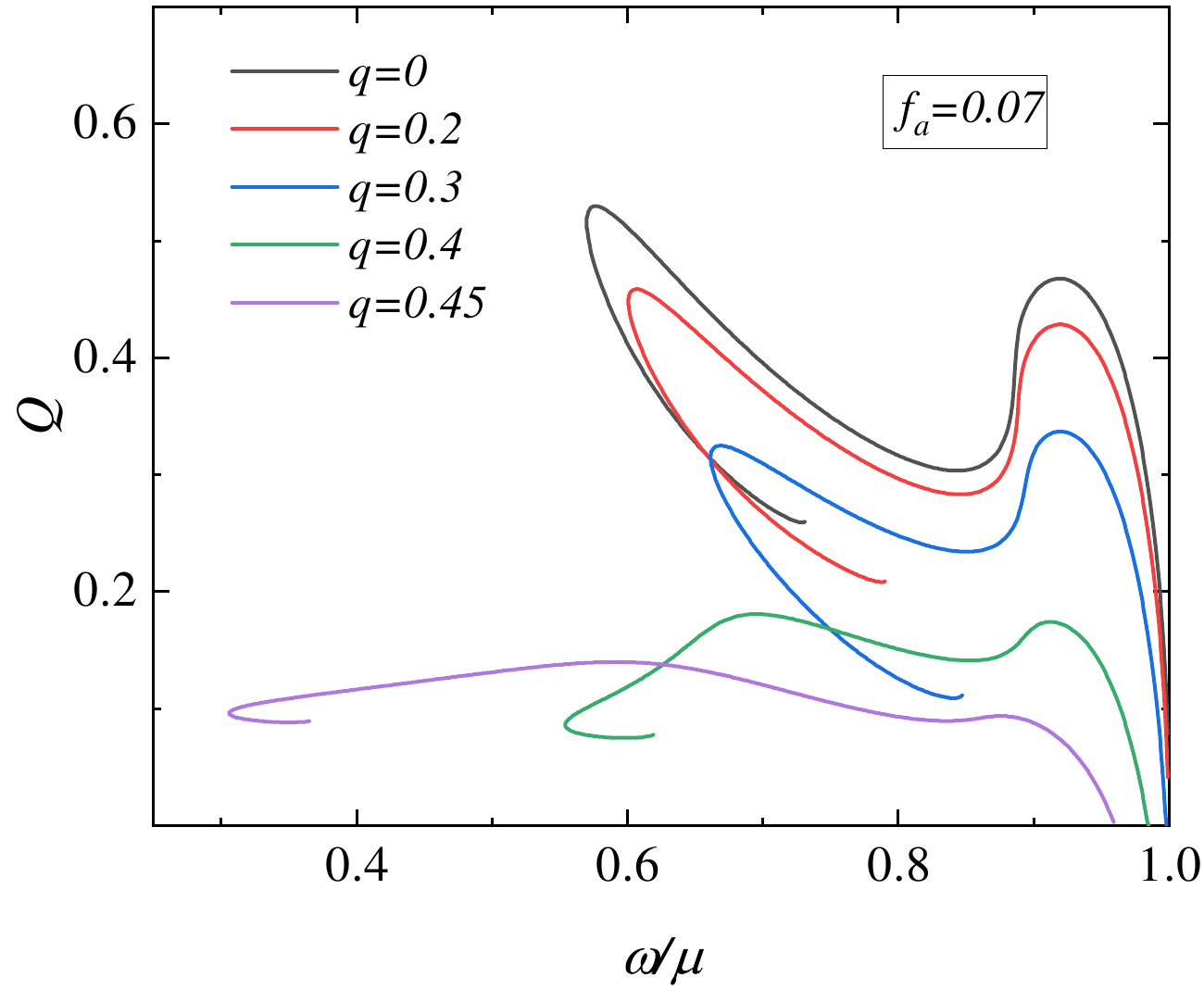}
   \includegraphics[width=0.45\textwidth]{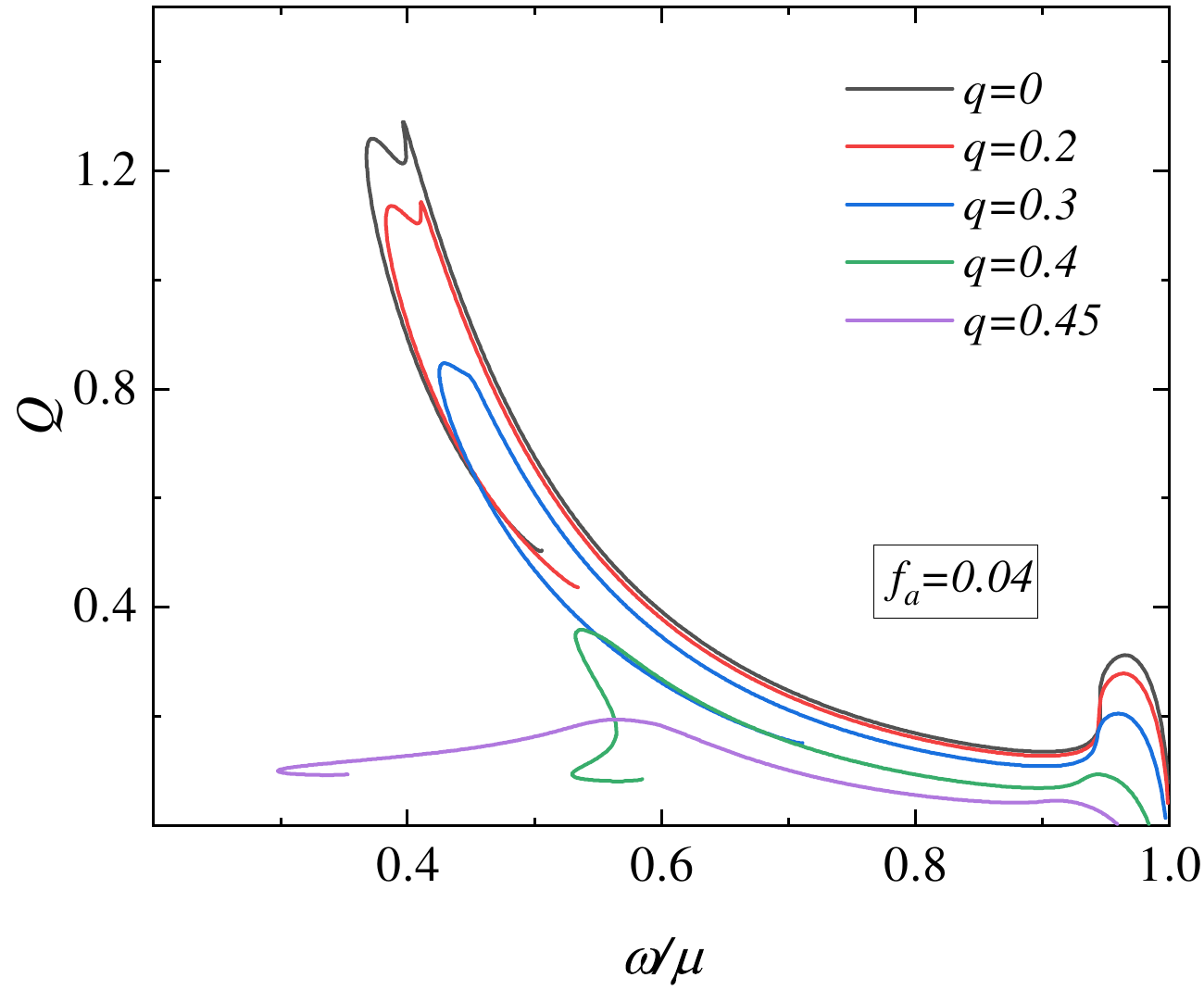}
   \includegraphics[width=0.45\textwidth]{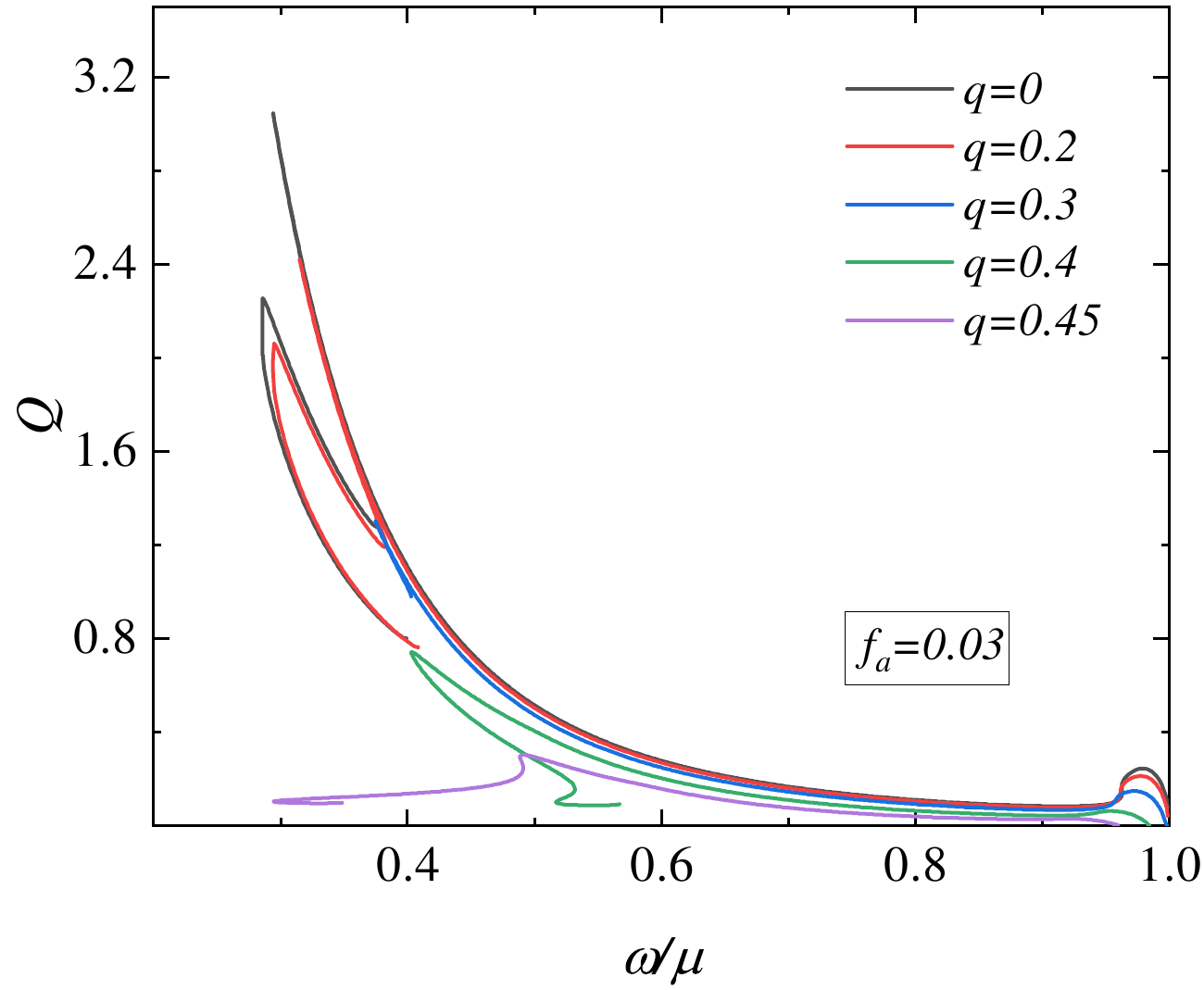}
   	\caption{The Noether charge $Q$ of the complex scalar field as a function of frequency $\omega$ for Hayward axion stars with low magnetic charge $q<q_c$ at decay constant $f_a=\{1,0.14,0.1,0.07,0.04,0.03\}$. Solid lines of the same color represent the same magnetic charge $q$, while the black solid line represents the solution for spherically symmetric axion stars without magnetic charge.}
   	\label{lowqQw}
   \end{figure} 
First, we give the Hayward axion star solutions for low magnetic charge $q < q_c$, where no solutions with frequency $\omega$ approaching zero can be found.
In Fig.~\ref{lowqmw}, the trend of the ADM mass $M$ with frequency $\omega$ for different magnetic charges $q=\{0,0.2,0.3,0.4,0.45\}$ when the decay constant $f_a=\{1,0.14,0.1,0.07,0.04,0.03\}$ is shown.
Solid lines of the same color represent the same magnetic charge $q$. The curve gradually unfolds as the magnetic charge $q$ increases.
For a fixed magnetic charge $q$, a decrease in the decay constant 
$f_a$ results in the mass $M$ of the Hayward axion star exhibiting multiple local maxima, which is similar to the spherically symmetric axion stars\cite{Guerra:2019srj}.
In Fig.~\ref{lowqQw}, we show the variation of the Noether charge $Q$ with frequency.
The trend of the Noether charge $Q$ with frequency is similar to the trend of mass $M$ with frequency.

\subsection{High magnetic charge ($q \geq q_c$)}

\begin{figure}[!h]
           	\centering
           \includegraphics[width=0.45\textwidth]{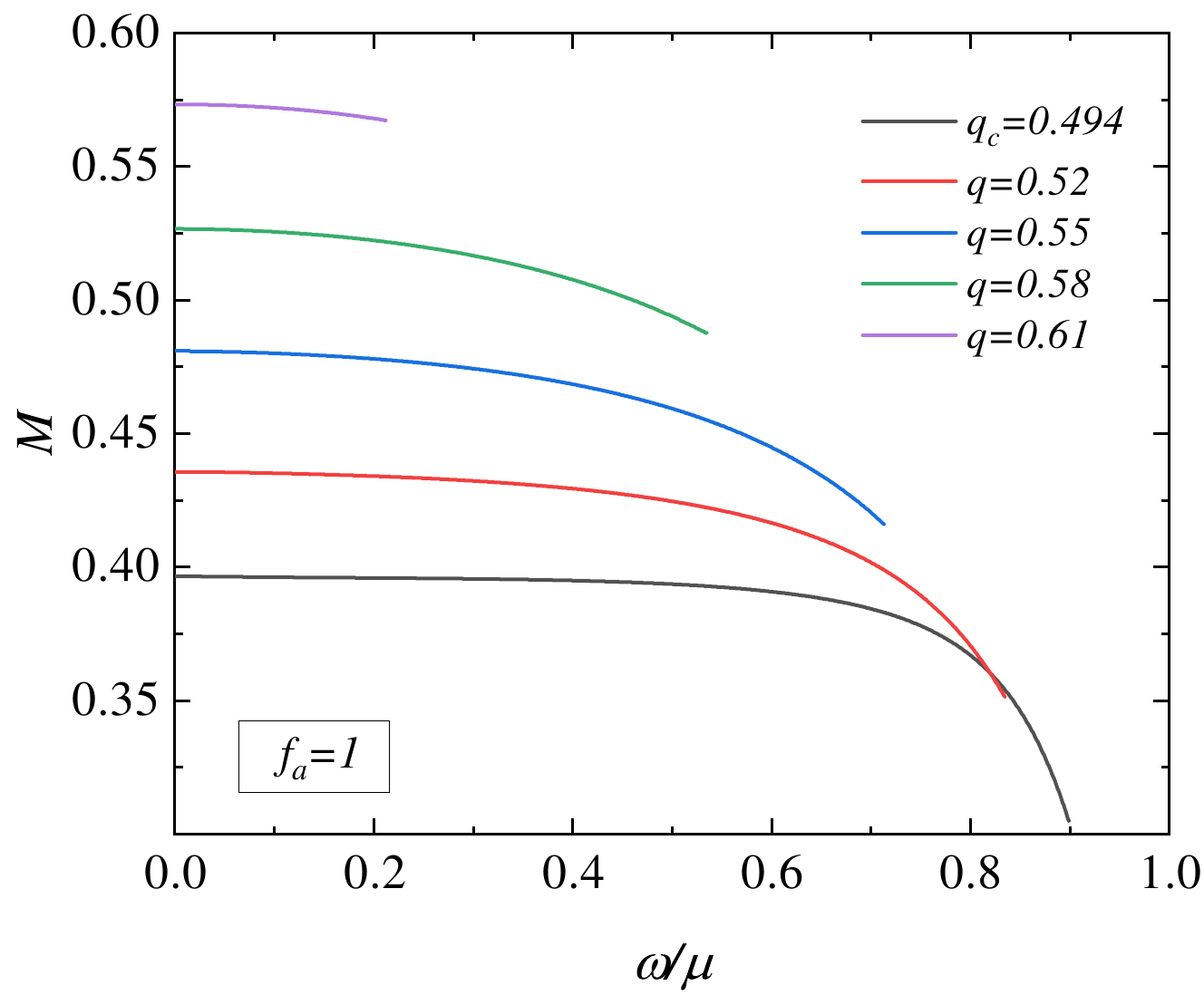}
           \includegraphics[width=0.45\textwidth]{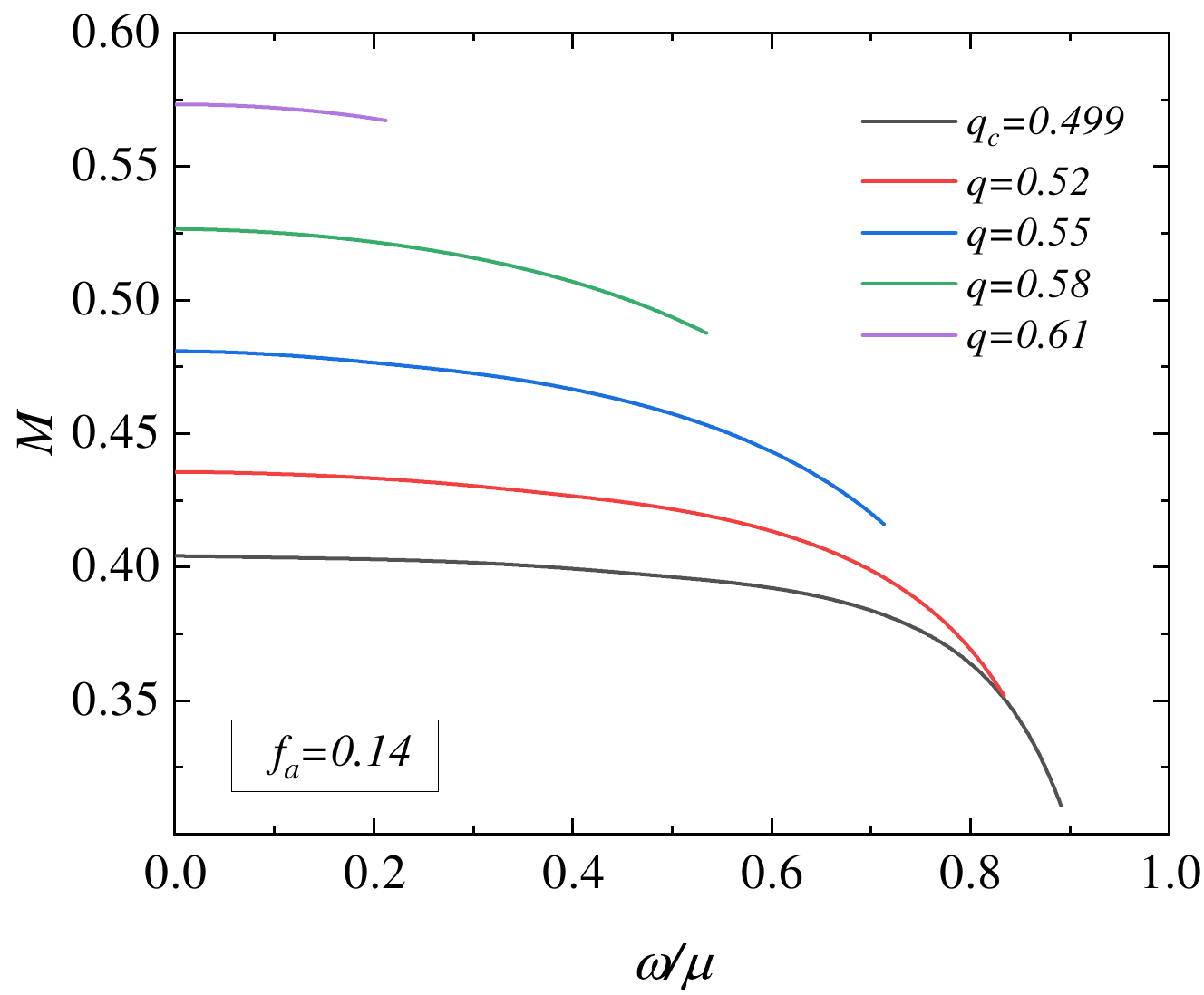}
           \includegraphics[width=0.45\textwidth]{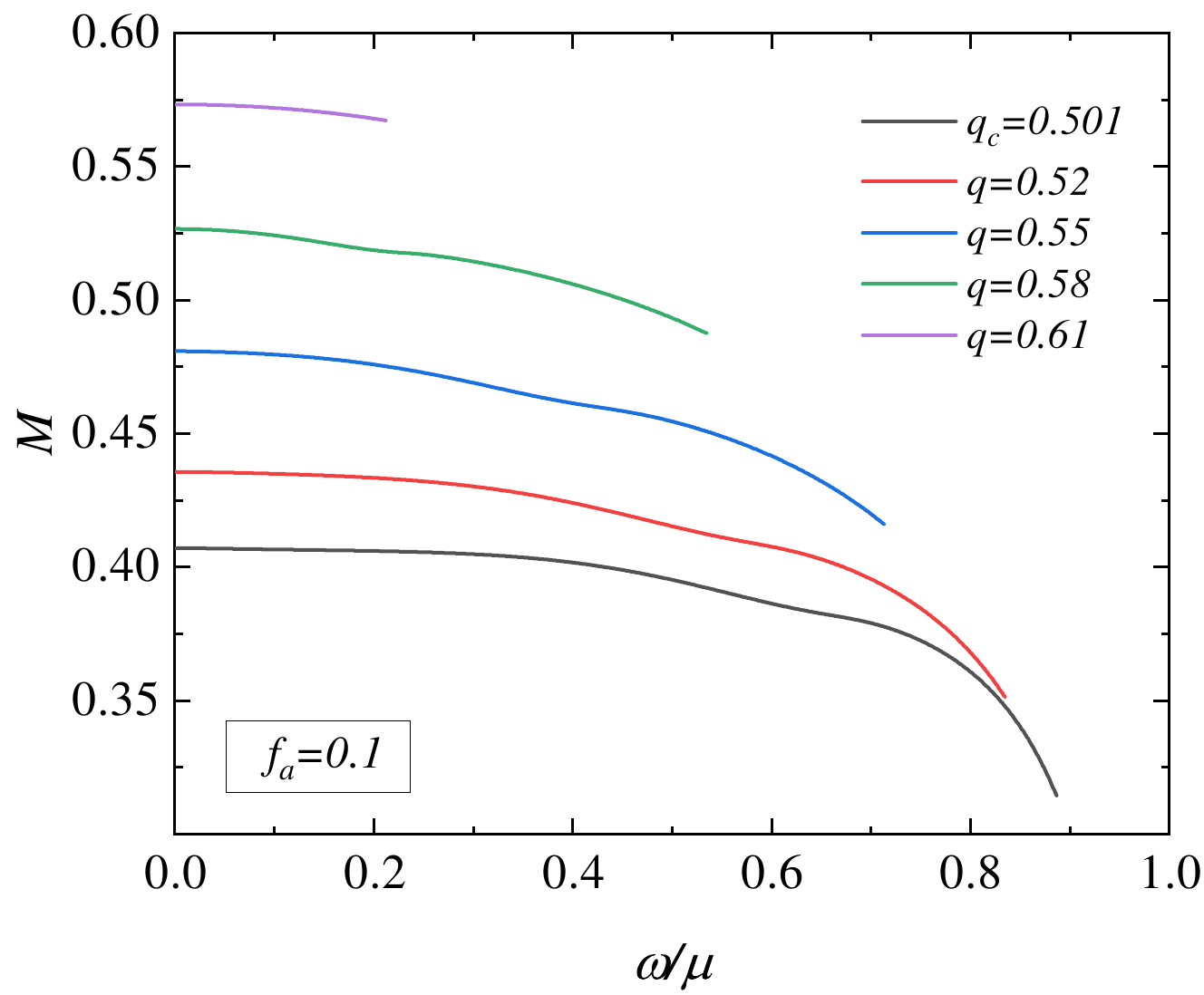}
           \includegraphics[width=0.45\textwidth]{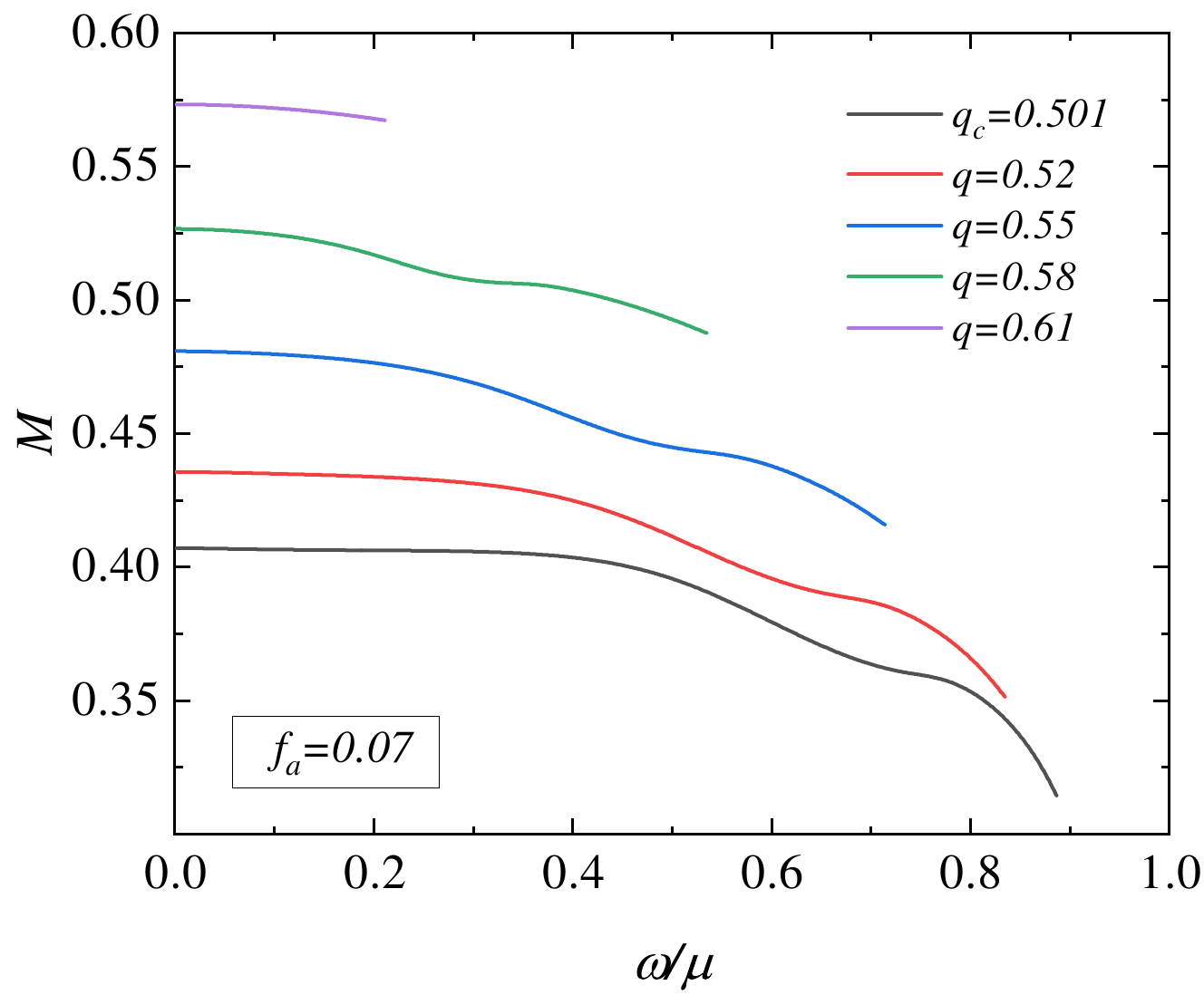}
           \includegraphics[width=0.45\textwidth]{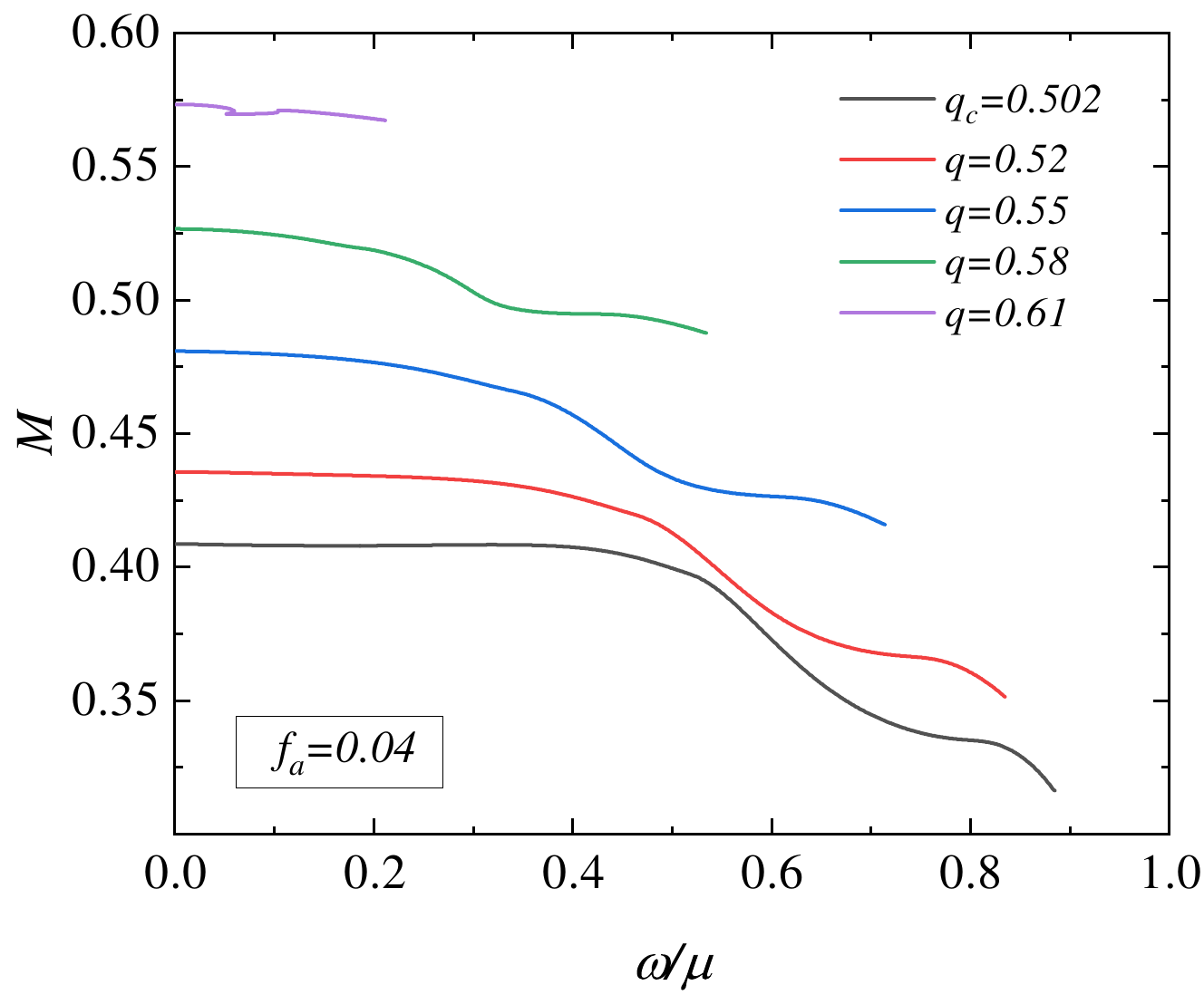}
           \includegraphics[width=0.45\textwidth]{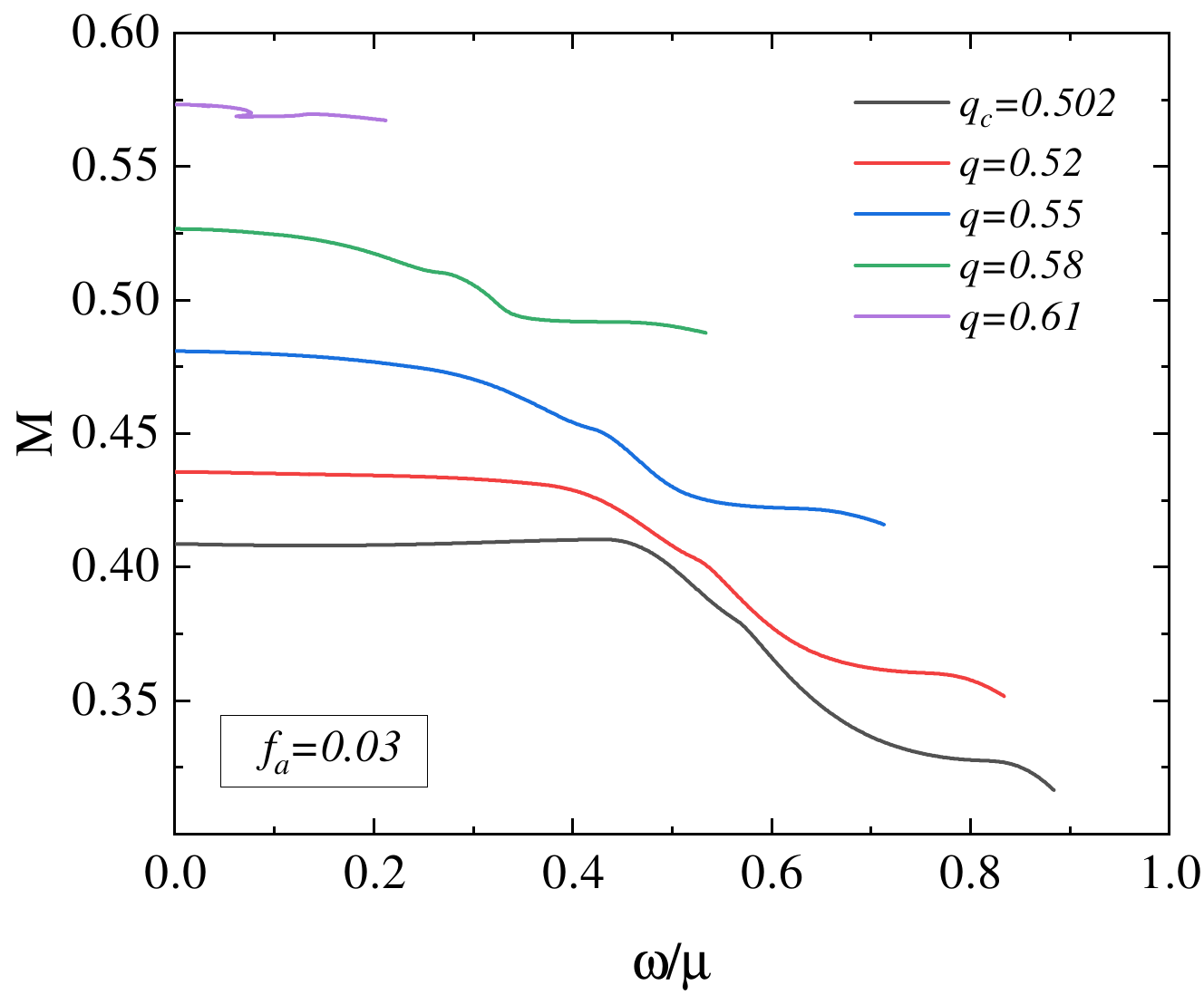}
           	\caption{The ADM mass $M$ of hayward axion stars as a function of frequency $\omega$ for the high magnetic charge $q\geq q_c$ at decay constant $f_a=\{1,0.14,0.1,0.07,0.04,0.03\}$. The black solid line represents the critical magnetic charge $q_c$.}
           	\label{highqmw}
           \end{figure}

Next, we give the Hayward axion star solutions for high magnetic charge $q\geq q_c$, where extreme solutions with frequency $\omega \to 0$ exist.
In Fig.~\ref{highqmw}, we show the relationship between the ADM mass $M$ and frequency $\omega$ for decay constant $f_a=\{1,0.14,0.1,0.07,0.04,0.03\}$. 
Different colored solid lines represent different magnetic charges $q$, and the black solid line represents the critical magnetic charge $q_c$ for the respective decay constant $f_a$. 
When the decay constant $f_a$ is held constant, the mass of extreme solutions increases as the magnetic charge $q$ increases.
When the magnetic charge $q$ is constant and the decay constant $f_a$ varies, the mass of extreme solutions remains unchanged. The masses $M$ for different magnetic charges $q$ at a frequency $\omega=0.0001$ for any given decay constant $f_a$ are given in Tab.~\ref{mmax}. This shows that the mass $M$ in extreme solutions is independent of the decay constant $f_a$ and only depends on the magnetic charge $q$.  However, when the frequency $\omega$ deviates from zero, the mass $M$ becomes dependent on the decay constant $f_a$.

       \begin{table}[htbp]
  \centering
  \begin{tabular}{|c|c|c|c|c|}
    \hline
   $q$ & 0.52 & 0.55 & 0.58 & 0.61\\
    \hline
   $M$ & 0.4957 & 0.4809 & 0.5268 & 0.5734\\
    \hline
  \end{tabular}
  \caption{The ADM mass $M$ of Hayward axion stars at a frequency $\omega = 0.0001$ for magnetic charges $q = \{0.52, 0.55, 0.58, 0.61\}$.}
    \label{mmax}
\end{table}

\begin{figure}[!htbp]
           	\centering
           	\includegraphics[width=0.45\textwidth]{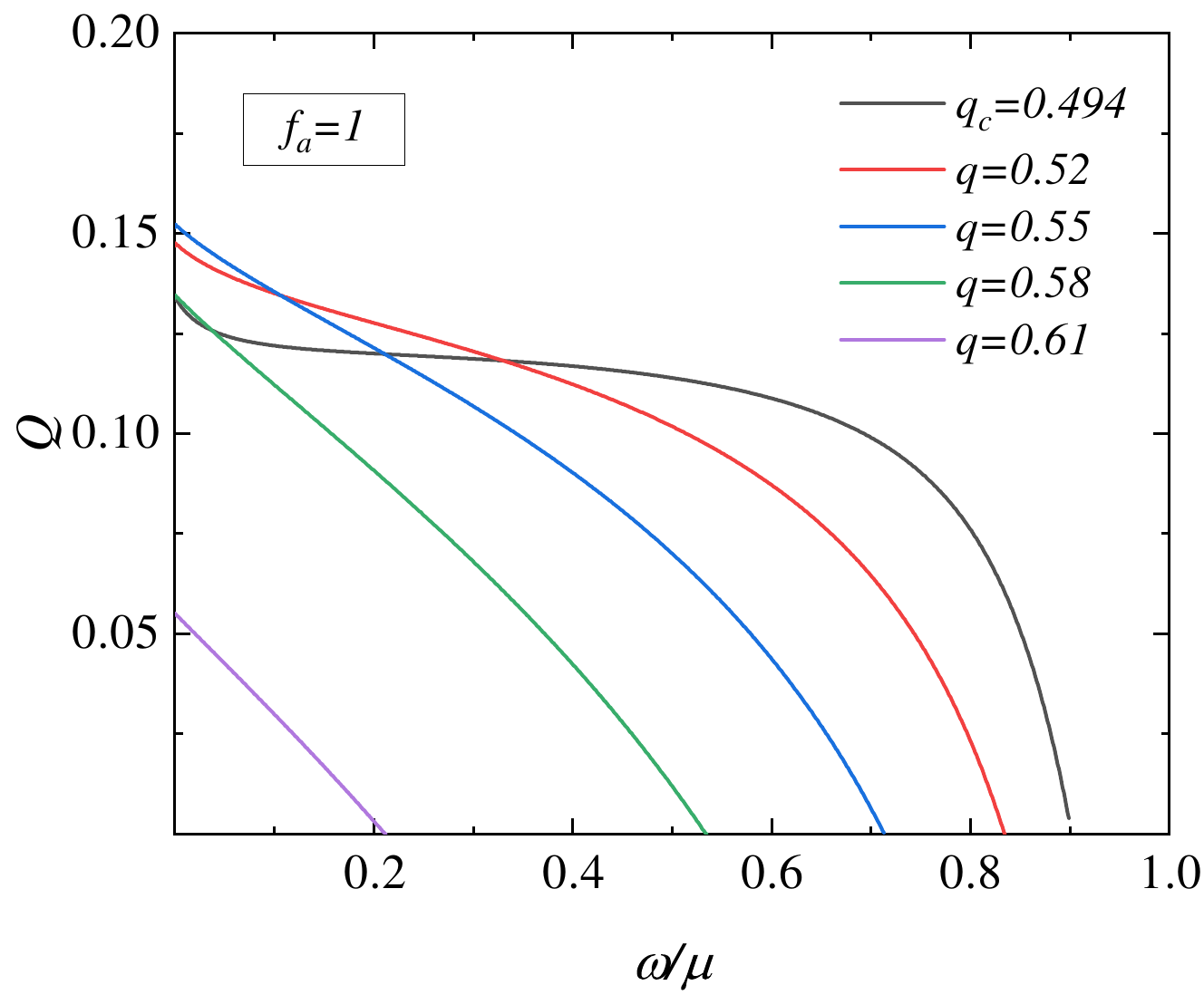}
           	\includegraphics[width=0.45\textwidth]{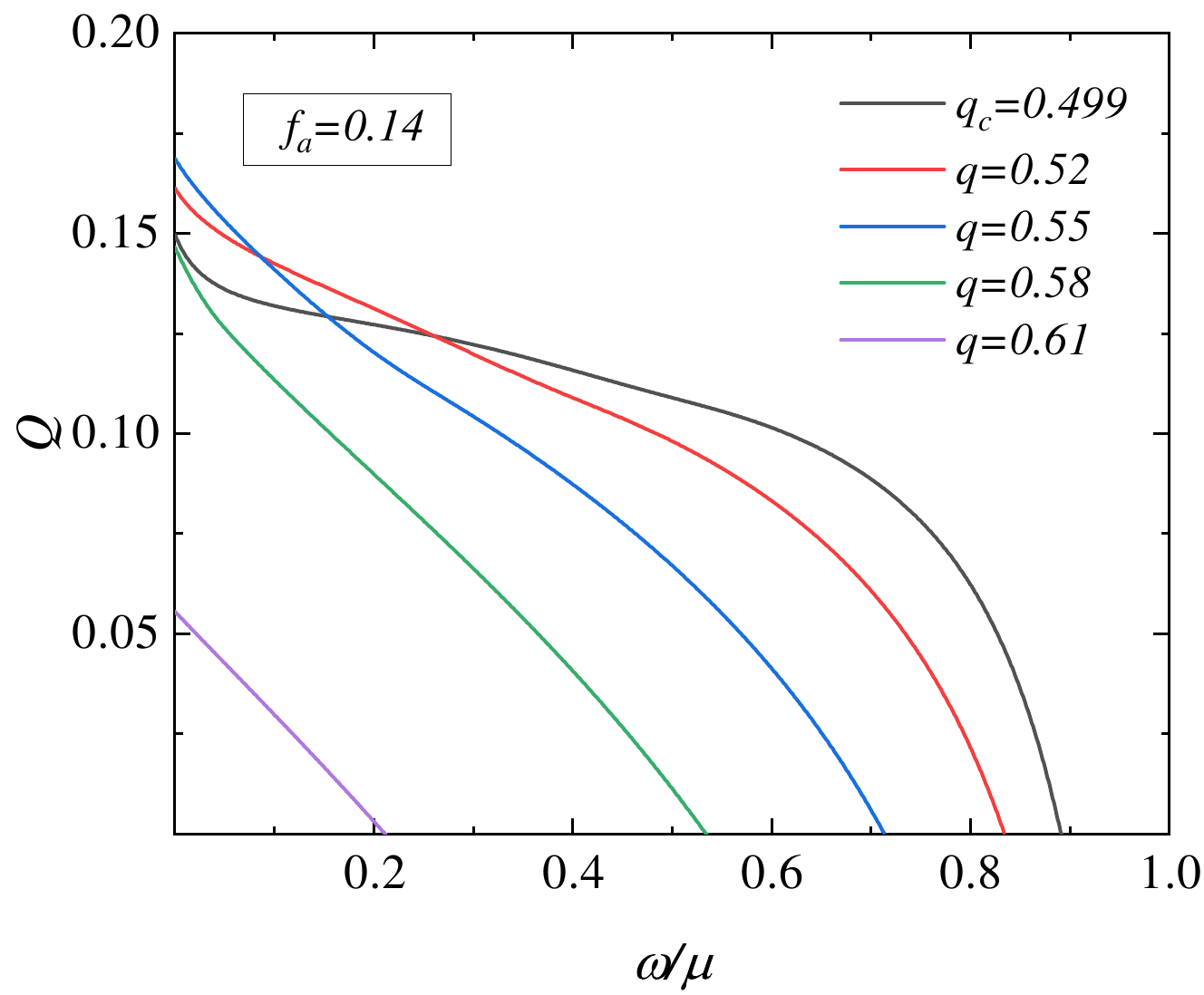}
           	\includegraphics[width=0.45\textwidth]{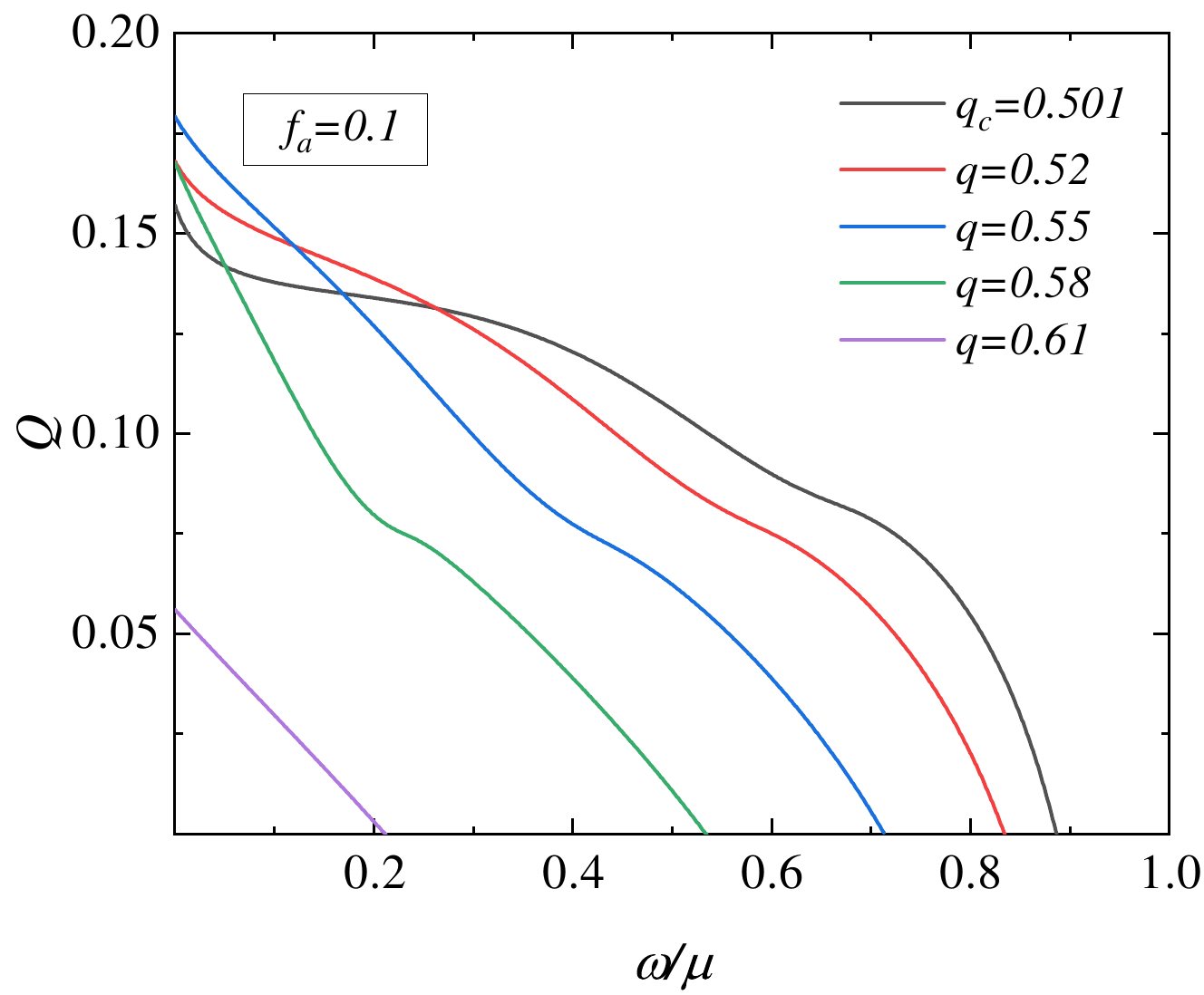}
           	\includegraphics[width=0.45\textwidth]{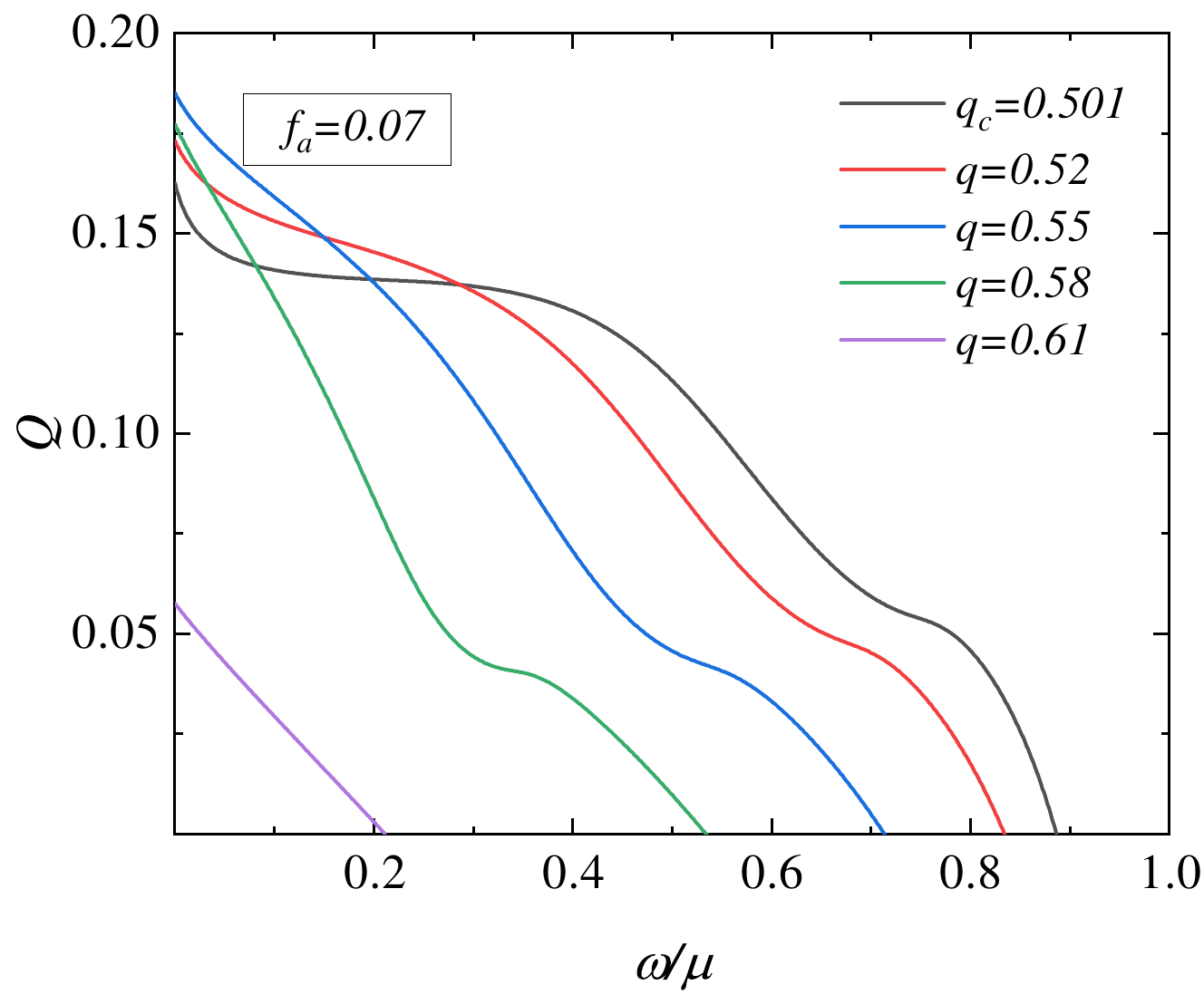}
           	\includegraphics[width=0.45\textwidth]{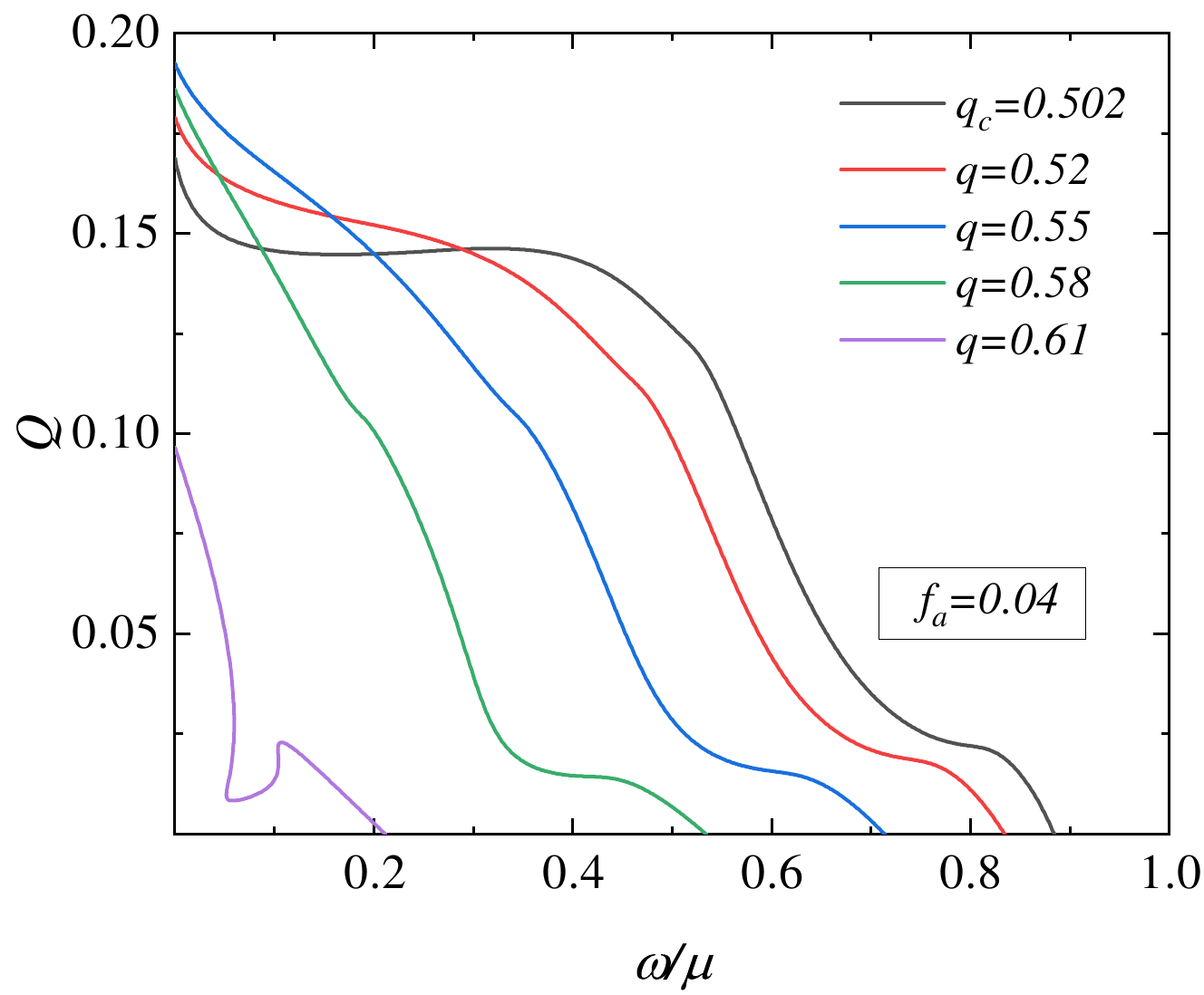}
           	\includegraphics[width=0.45\textwidth]{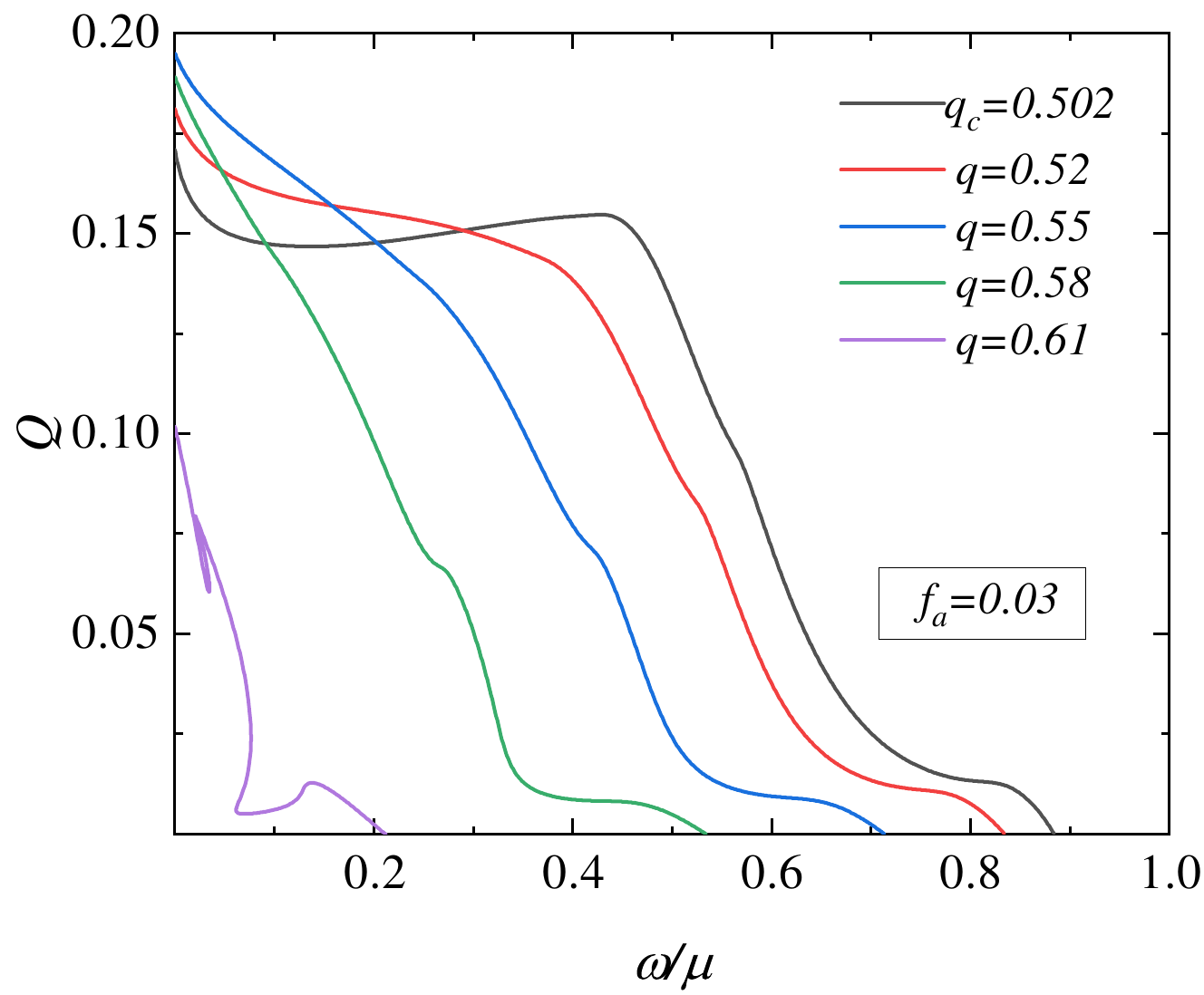}
           	\caption{The Noether  charge $Q$ of the complex scalar field as a function of frequency $\omega$ for the high magnetic charge $q\geq q_c$ with decay constant $f_a=\{1,0.14,0.1,0.07,0.04,0.03\}$. The black solid line represents the critical magnetic charge $q_c$.}
           	\label{highqQw}
           \end{figure} 

In Fig.~\ref{highqQw}, we illustrate the relationship between the Noether charge $Q$ and the frequency $\omega$ for high magnetic charge $q\geq q_c$. The solid lines of different colors represent different magnetic charges $q$. For a fixed decay constant $f_a$, the domain of solutions becomes smaller as the magnetic charge $q$ increases. Notably, unlike the mass,  the Noether charge $Q$ is significantly influenced by the decay constant $f_a$ in the extreme solutions.

\begin{figure}[htbp]
  \centering
  \includegraphics[width=0.45\textwidth]{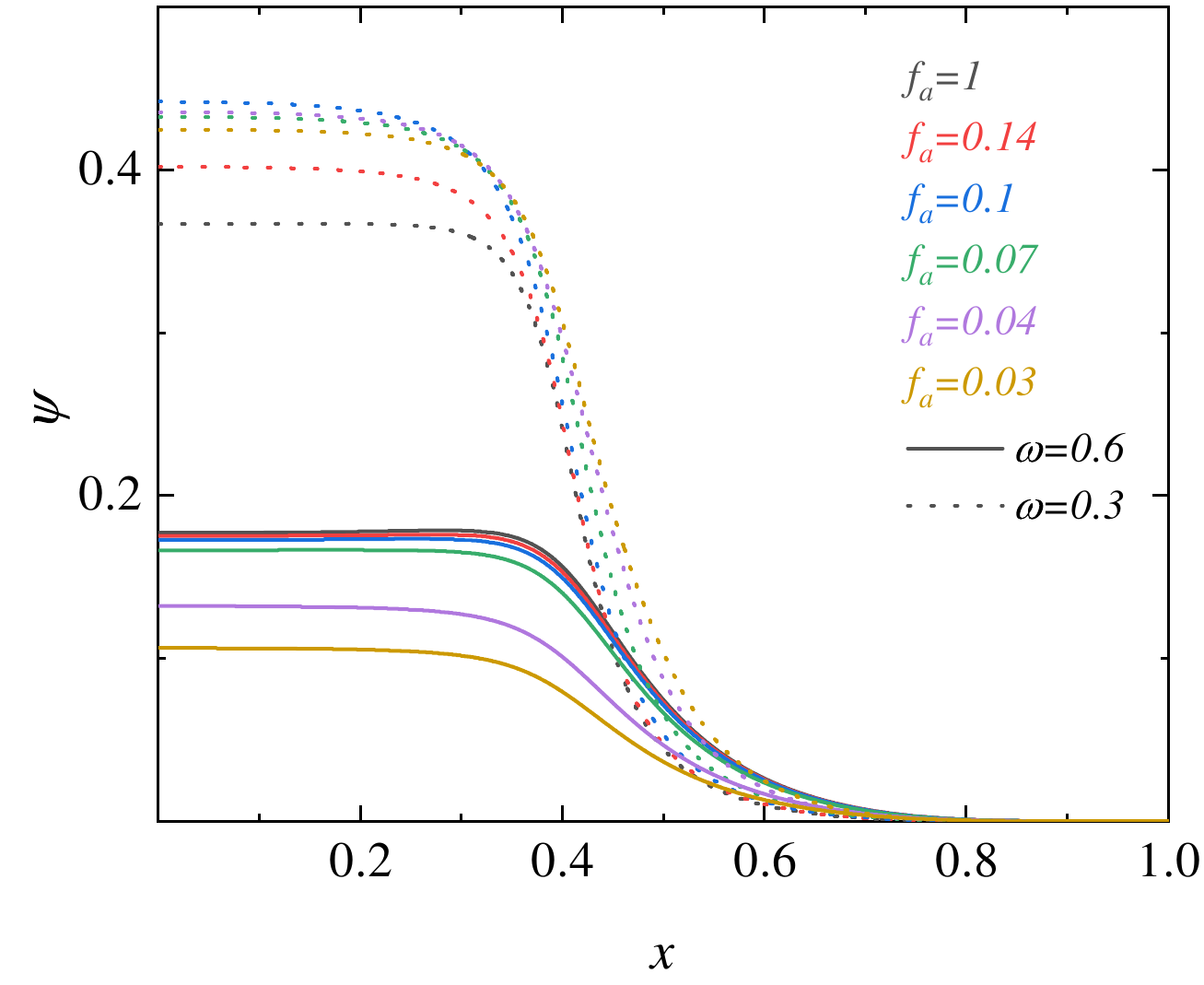}
  \includegraphics[width=0.45\textwidth]{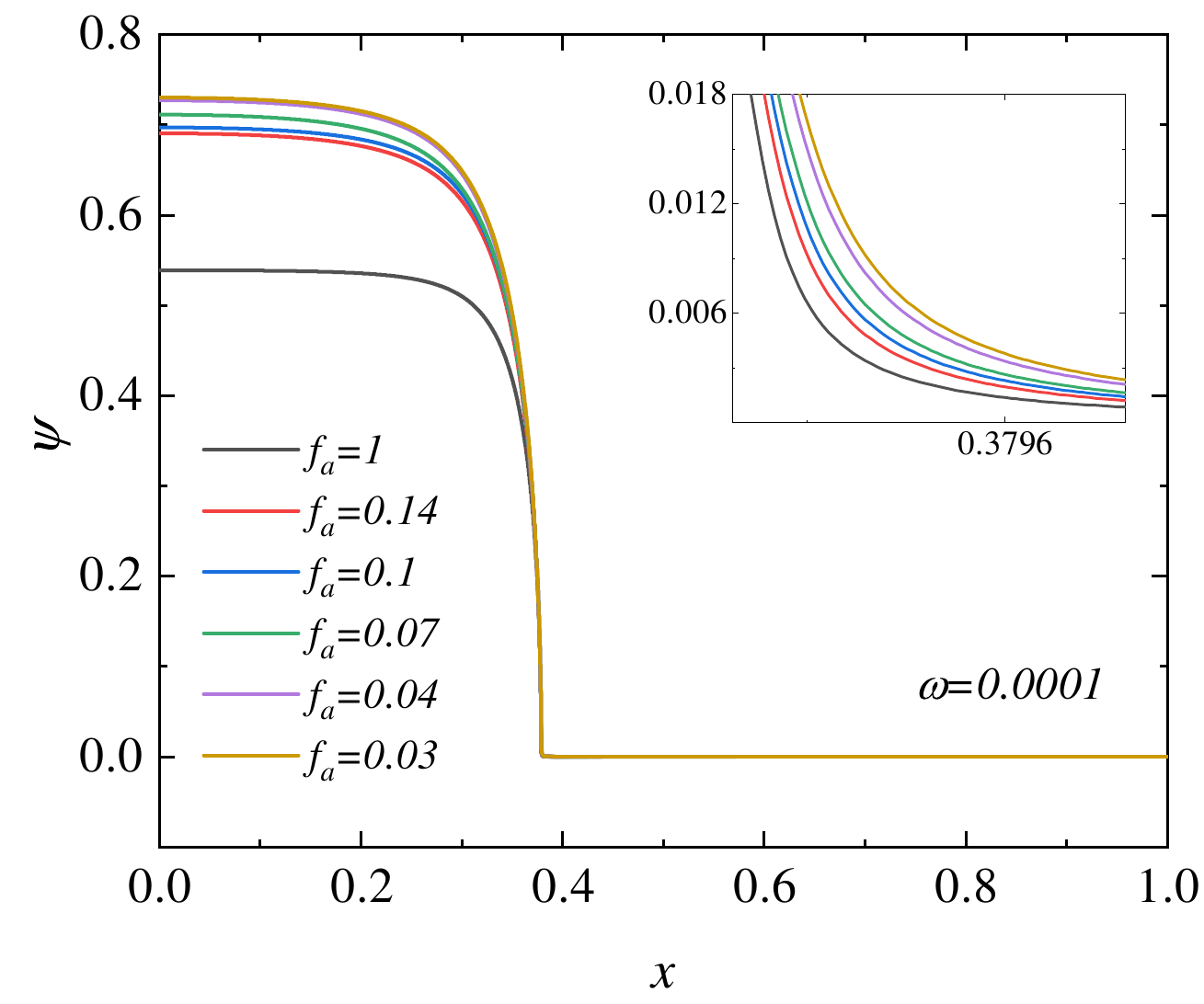}
  \includegraphics[width=0.45\textwidth]{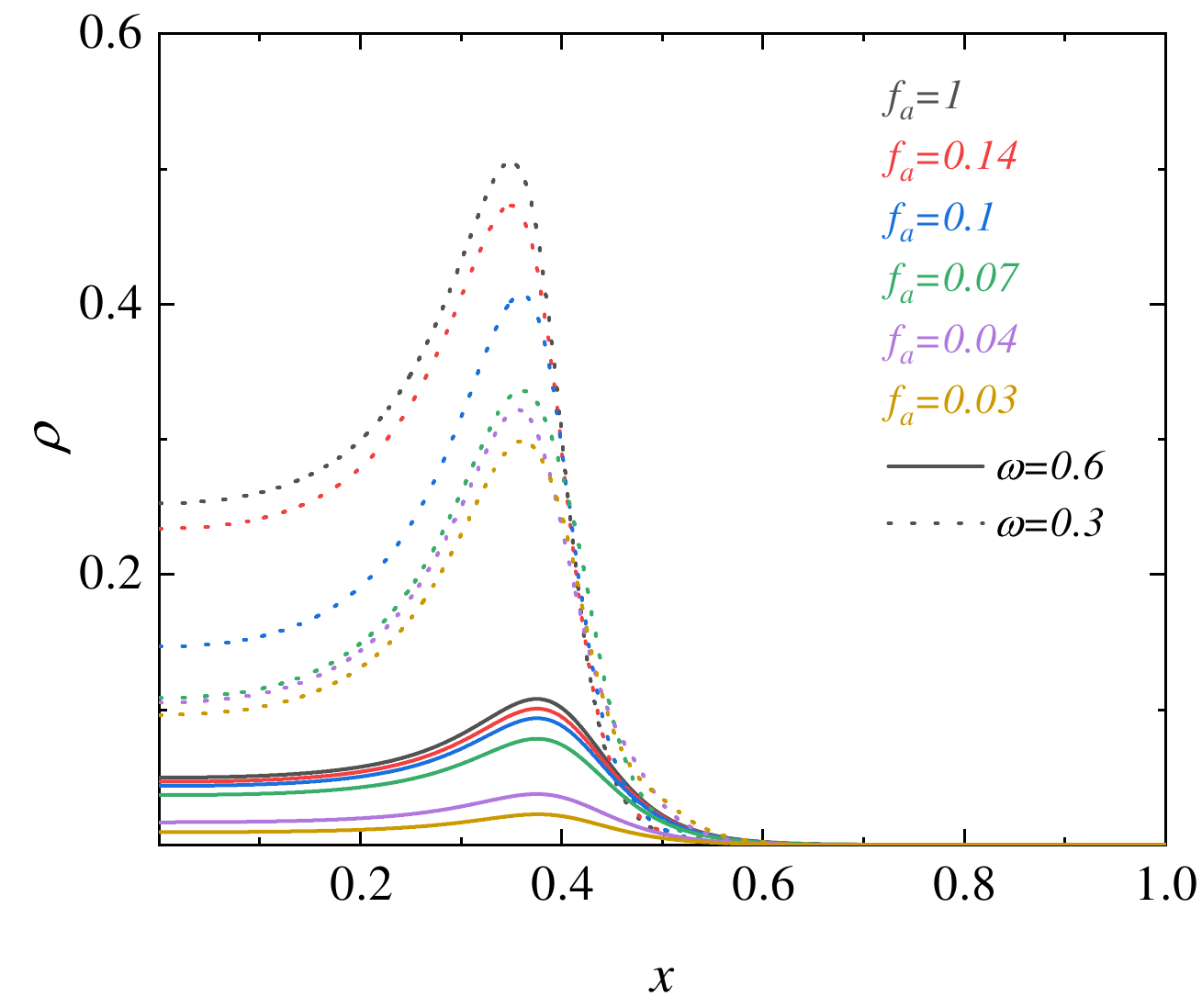}
  \includegraphics[width=0.45\textwidth]{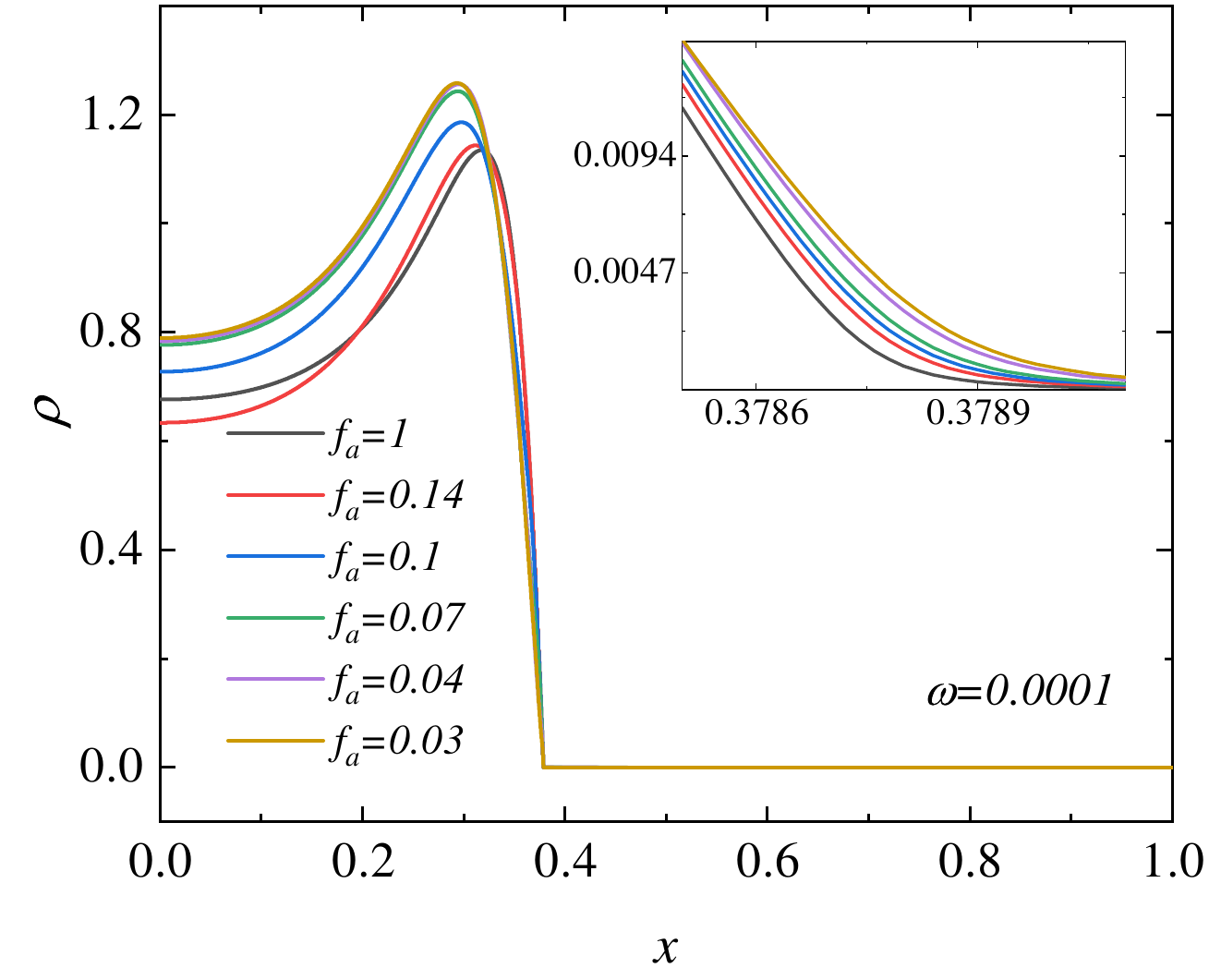}
  \caption{ The distribution of field functions $\psi$ and energy densities $\rho$ along radial coordinates $r$ for different decay constants $f_a$ with magnetic charge $q=0.55$, $s=0.2$ at frequencies $\omega=0.6,0.3$ (left) and $\omega=0.0001$ (right). The top panels plot the distribution of the field function $\psi$ along the radial coordinate $r$, and the bottom panels display the distribution of the energy density $\rho$ along the radial coordinate $r$.}
  \label{distribution1}
\end{figure}

To explore the behavior of the extreme solutions, 
  in Fig.~\ref{distribution1} we depict the radial distribution of the scalar field $\psi$ and energy density $\rho$ of Hayward axion stars for different decay constants $f_a$ at frequencies $\omega = \{0.6, 0.3, 0.0001\}$, with magnetic charge $q = 0.55$ and $s = 0.2$.
 Lines of the same color represent the same decay constant $f_a$.
 In the left panels, the solid and dashed lines respectively correspond to  $\omega=0.6$ and $\omega=0.3$. 
  The right panels show the solutions for $\omega =0.0001 $, with an inset displaying the details of the curves.
The top panels are the distribution of the scalar field $\psi$ with radial coordinates, and we can find that at $\omega=0.6$ and $\omega=0.3$, the field function is mainly distributed within a certain radius and begins to decrease exponentially beyond this radius. At $\omega=0.0001$, the scalar field declines rapidly, forming a very steep surface beyond a certain radius, which we refer to as the critical radius.
 The bottom panels are the variation of the energy density $\rho$ of the scalar field with radial coordinates. The energy density initially increases and then decreases along the radial coordinate. 
 At $\omega=0.0001$, the energy density drops sharply at the critical radius. 
 Variations in the decay constants $f_a$ lead to differences in the energy density distribution within the critical radius, but outside this radius, the energy density approaches zero.

\begin{figure}[htbp]
  \centering
   \includegraphics[width=0.45\textwidth]{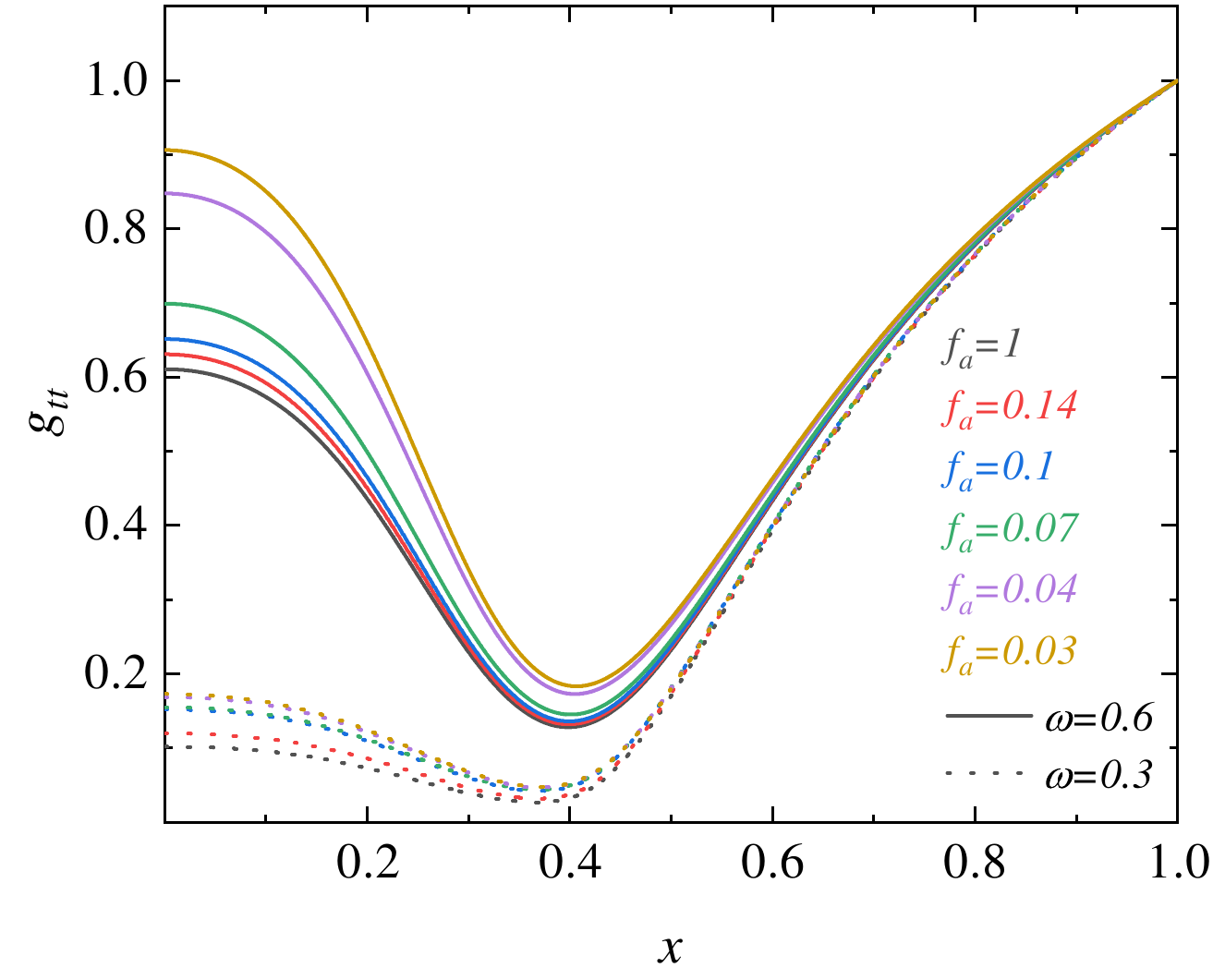}
  \includegraphics[width=0.45\textwidth]{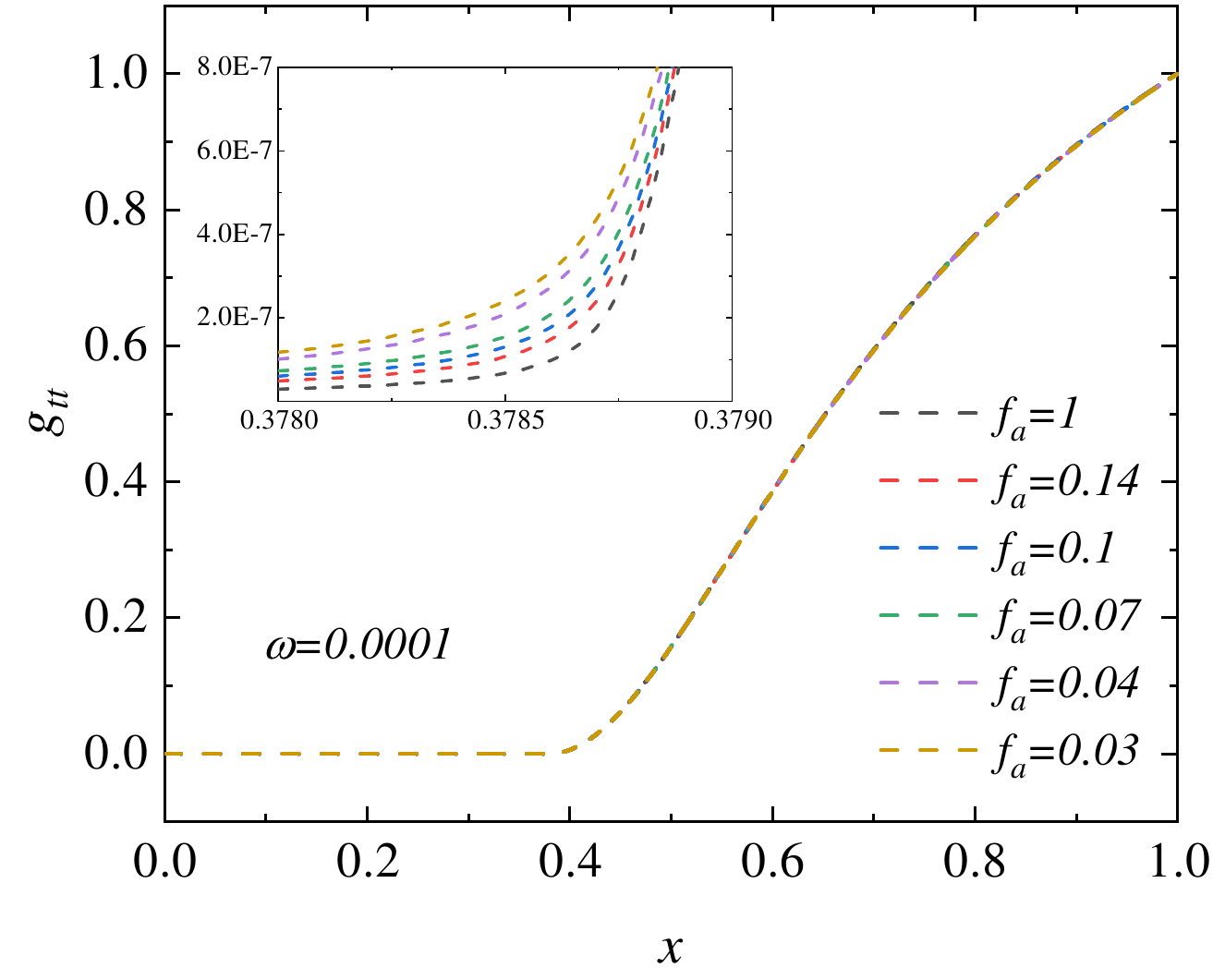}
  \includegraphics[width=0.45\textwidth]{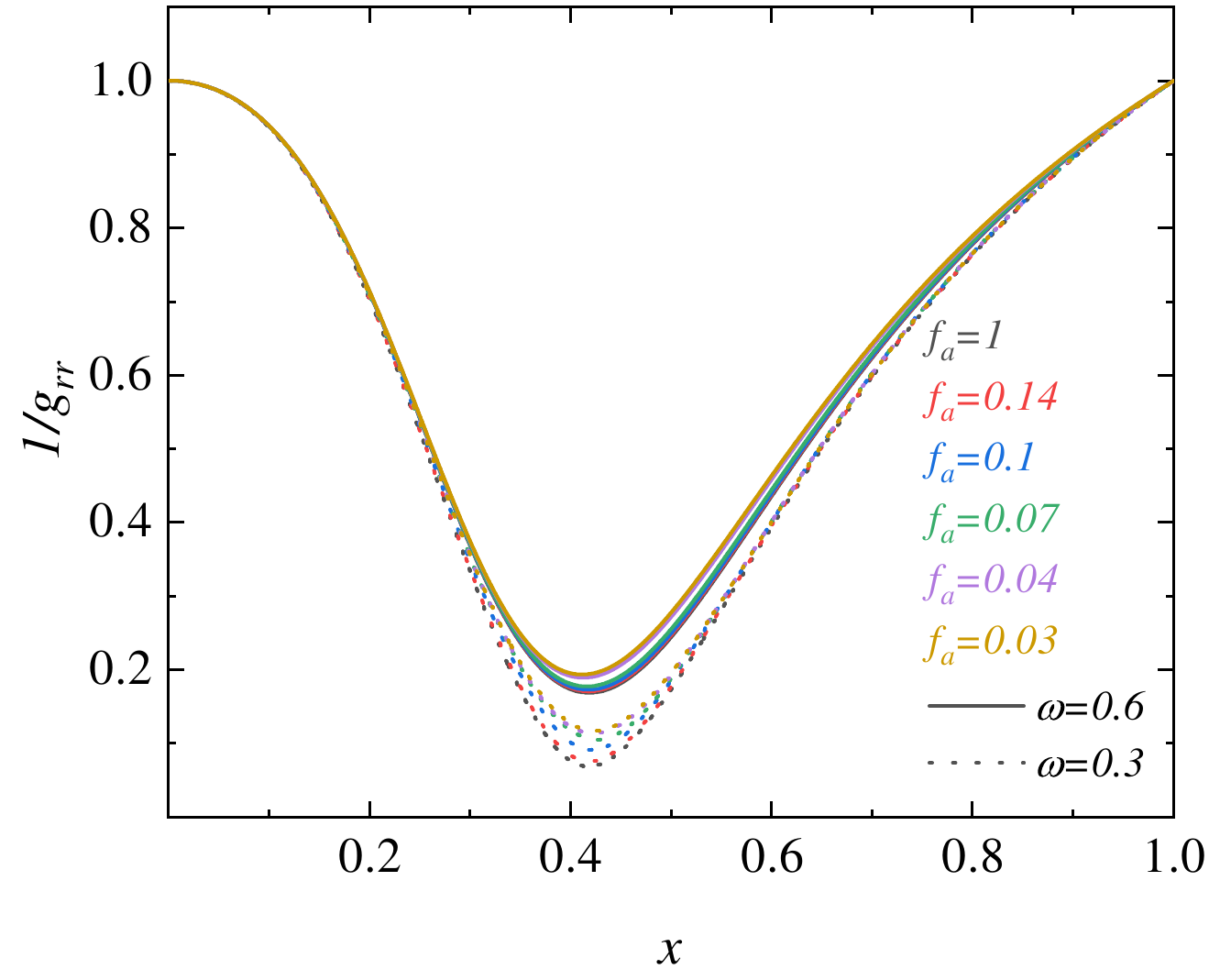}
  \includegraphics[width=0.45\textwidth]{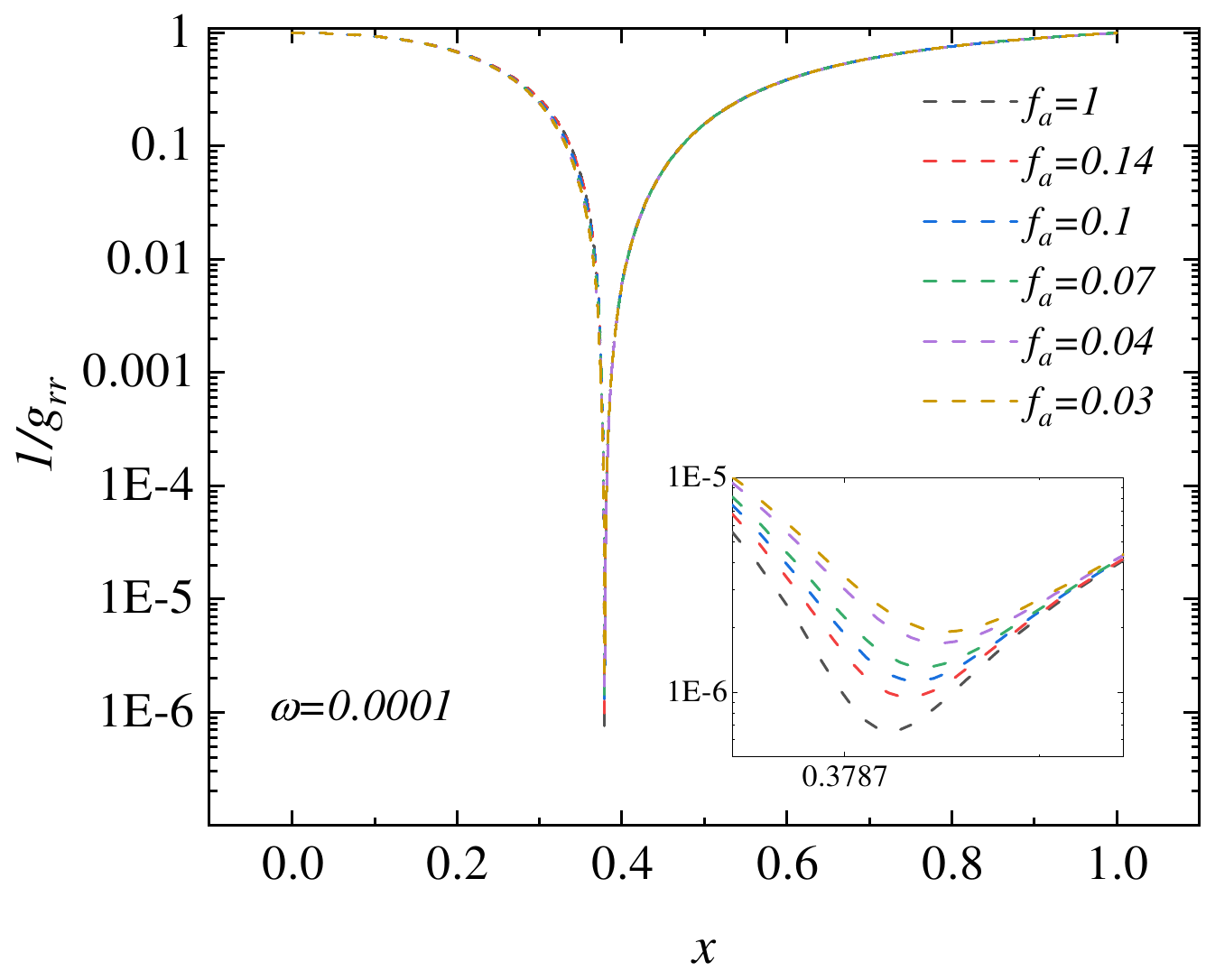}
  \caption{The distribution of the metric functions along radial coordinates $r$ for different decay constants $f_a$ for different decay constants $f_a$ with magnetic charge $q=0.55$, $s=0.2$ at frequencies $\omega=0.6,0.3$ (left) and $\omega=0.0001$ (right). The top panels show the distribution of the time component of the metric $g_{tt}$ along the radial coordinate $r$, and the distribution of the inverse of the spatial component of the metric $1/g_{rr}$ along the radial coordinate $r$ is plotted at the bottom.}
  \label{distribution2}
\end{figure}

In Fig.~\ref{distribution2}, we give the distribution of the metric components of the Hayward axion stars with radial coordinates at frequencies $\omega=\{0.6,0.3,0.0001\}$ for $q=0.55$ and $s=0.2$. In the left panels, the solid and dashed lines correspond to $\omega=0.6$ and $\omega=0.3$, respectively. The solutions for $\omega =0.0001 $ are shown in the right panels, with an inset displaying the details of the curves. The top panels display the variation of the metric component $g_{tt}$ with radial coordinates and the inset highlighting $g_{tt}$ near the critical radius. At  $\omega=0.0001$, $g_{tt}$ approaches zero at the critical radius influenced by the matter field.
The bottom panels illustrate the distribution of the inverse of the spatial component of the metric $1/g_{rr}$ along the radial coordinates.
The vertical axis on the right of the bottom panels is in a logarithmic scale for better observation. 
At $\omega=0.0001$, $1/g_{rr}$ tends to zero near the critical position, nearly forming the event horizon, which we call the critical horizon. 
Within the critical horizon, $g_{tt}$ is nearly zero, suggesting that time almost stops within this region.
To a distant observer, matter approaching the surface of the star appears to stop near the critical horizon, thus these solutions are termed Hayward axion frozen stars. 
Furthermore, when $\omega \to 0$ the curves of the metric for various decay constants $f_a$ almost coincide, indicating that the metric components of Hayward axion frozen stars are insensitive to the values of $f_a$.

\begin{figure}[!htbp]
  \centering
  \includegraphics[width=0.5\textwidth]{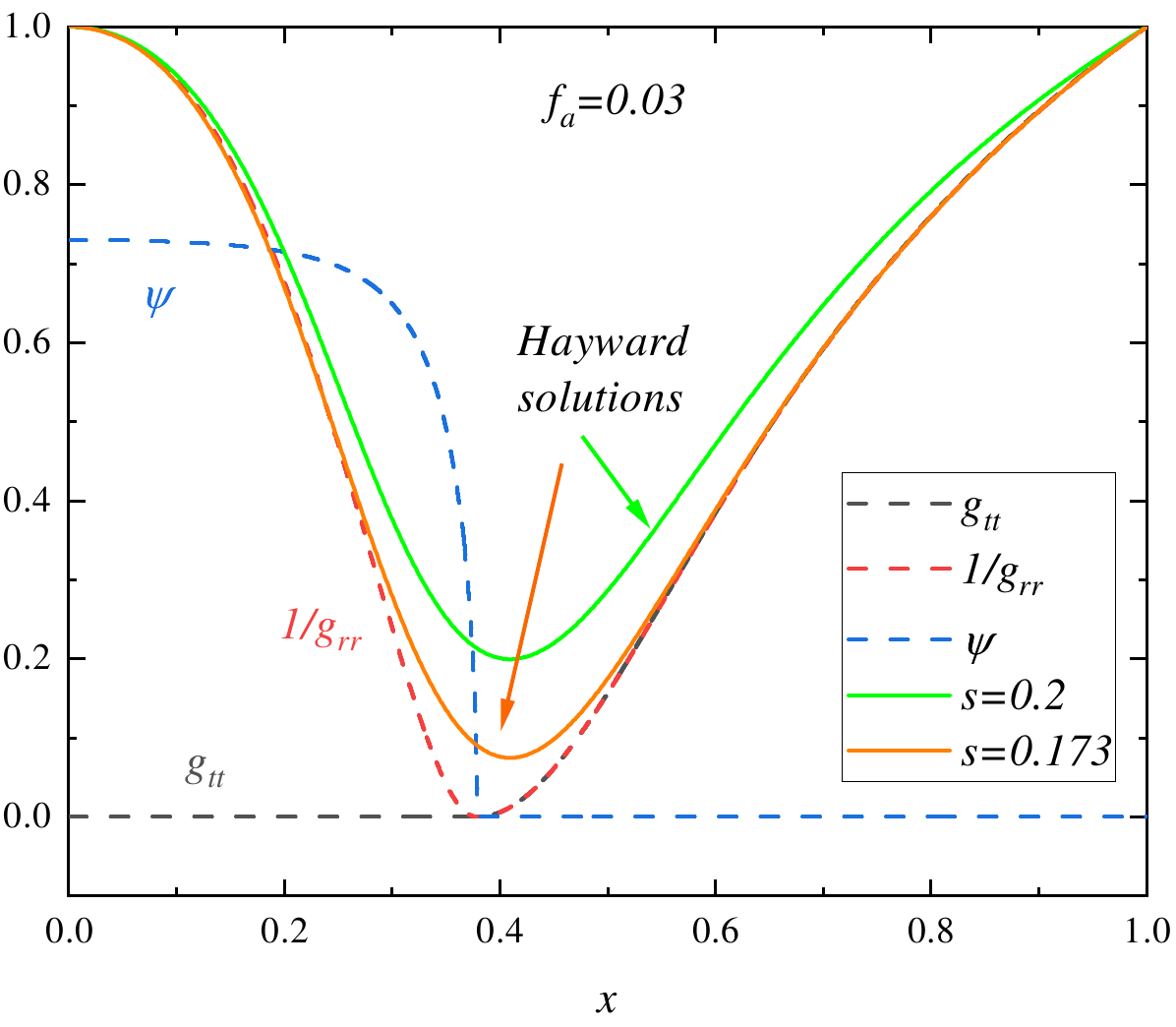}
  \caption{The distribution of solutions along radial coordinates. The green and orange lines respectively represent single Hayward solutions at  $s=0.2$ and $s=0.173$. 
   The dashed lines are the Hayward axion frozen star solutions at frequency $\omega=0.0001$.
 The blue, red, and black dashed lines represent the field function $\psi$, the inverse of the spatial component of the metric $1/g_{rr}$, and the time component of the metric $g_{tt}$, respectively. All lines have the same value with decay constant 
$f_a=0.03$ and magnetic charge $q=0.55$.}
  \label{hay}
\end{figure}

To further understand the solutions of Hayward axion frozen stars,  we compare the Hayward axion frozen star solutions with the single Hayward solutions in Fig.~\ref{hay} for a decay constant $f_a = 0.03$, magnetic charge $q = 0.55$, and frequency $\omega = 0.0001$.   Under the Hayward metric, $s = \frac{q^3}{2M}$, and when $q = 0.55$, we derive $s = 0.173$.
In Fig.~\ref{hay}, the dashed lines depict the solutions of the Hayward axion frozen star varying with the radial coordinate. 
 The green and orange solid lines represent the single Hayward solutions for $s = 0.2$ and $s = 0.173$, respectively.          
The orange and red curves have the same mass.
 In the absence of the axion field, the single Hayward solution lacks the event horizon.         
However, with the axion field present, the Hayward axion frozen star solution features a critical horizon.         
 For a distant observer, the Hayward axion frozen star solution is indistinguishable from the single Hayward solution.

\subsection{Light ring}

 General relativity predicts that light will bend when passing through a strong gravitational field, and it can even travel along closed orbits~\cite{Einstein:1916vd}. For sufficiently compact objects, photons can form closed circular orbits under strong gravitational influence, which are known as light rings ~\cite{Cardoso:2014sna}. Next, we will investigate the presence of light rings around the Hayward axion star. In a gravitational field, photons move along null geodesics~\cite{Thorne:1973}
\begin{equation}
    g_{\mu \nu} \Dot{x}^{\mu}\Dot{x}^{\nu}=0\ ,
    \end{equation}
where $x^{\mu}=(t,r,\theta,\phi)$, and the dot denotes derivative with respect to the affine parameter $\lambda$ of the geodesic.
For spherically symmetric static spacetime, the motion of particles can be confined to the equatorial plane $(\theta=\pi/2)$. Due to the lack of explicit dependence of the metric components on $t$ and $\phi$, the energy $E$ and angular momentum $L$ of the photon are conserved. The energy per unit mass is defined as $E=-g_{tt}\dot{t}$ and the angular momentum per unit mass as $L=r^2 \dot{\varphi}$~\cite{chandrasekhar1998mathematical}. These conserved quantities, $E$ and $L$, can be substituted into the geodesic equation for photons to yield 
\begin{equation}
    \Dot{r}^2 +\frac{L^2}{g_{tt} g_{rr}}\left(\frac{1}{b^2}+\frac{g_{tt}}{r^2}\right)=0\ ,
\end{equation}
where $b$ denotes the impact parameter, defined as $b\equiv\frac{L}{E}$. Substituting the metric ansatz into the above equation gives
\begin{equation}
\Dot{r}^2+\frac{L^2}{e^u  e^v}\left(\frac{e^u}{r^2}-\frac{1}{b^2}\right)=0\ .
\end{equation}
We define the effective potential as
\begin{equation}
    V_{eff}=\frac{e^u}{r^2}\ .
    \label{effective}
\end{equation}
Assuming a null particle originating from a distant location arrives at $R_{lr}$. If this particle satisfies the condition $\frac{1}{b_1^2} = \frac{e^u(R_{lr})}{R_{lr}^2}$, its radial velocity will decrease to zero, causing it to remain stationary at that position.
The orbit of the null particle at $r = R_{lr}$  is referred to as light rings~\cite{Cardoso:2021sip}. 
 By setting the derivative of the effective potential with respect to the radial coordinate to zero
  \begin{equation}
    \frac{dV_{eff}}{dr}\bigg|_{R_{lr}} =0\ ,
     \label{lightring}
\end{equation}
we can determine the location of the light ring $R_{lr}$. The stability of the light ring is determined by the second derivative of the effective potential at the light ring $R_{lr}$~\cite{Cunha:2017qtt, Cunha:2022gde}.
If $V^{''}_{eff}(R_{lr})<0$, the effective potential exhibits a maximum at this point, indicating that the light ring is unstable. 
 Conversely, if $V^{''}_{eff}(R_{lr}) > 0$, the effective potential has a minimum, indicating the stable light ring.

\begin{figure}[!htbp]
  \centering
  \includegraphics[width=0.45\textwidth]{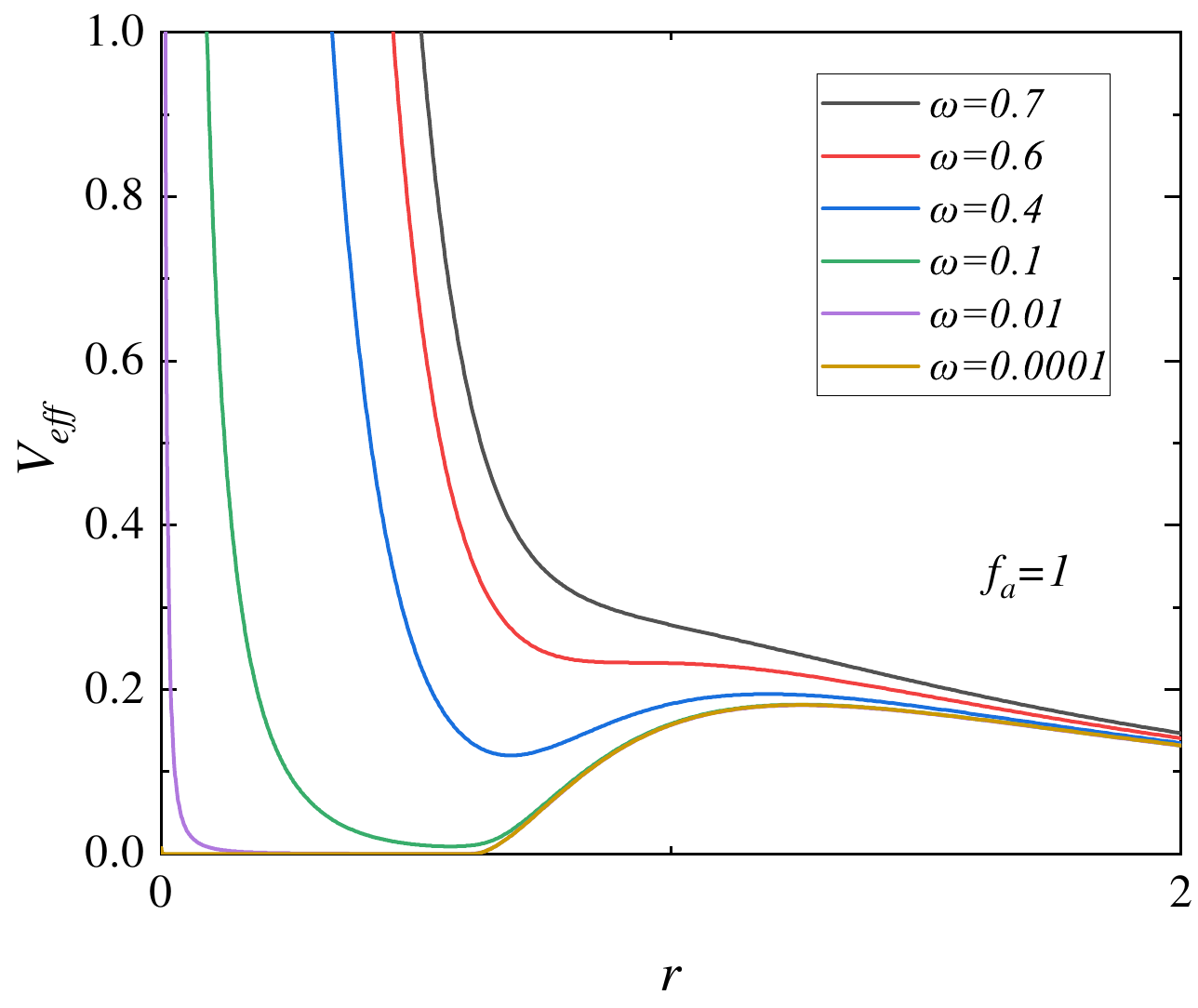}
  \includegraphics[width=0.45\textwidth]{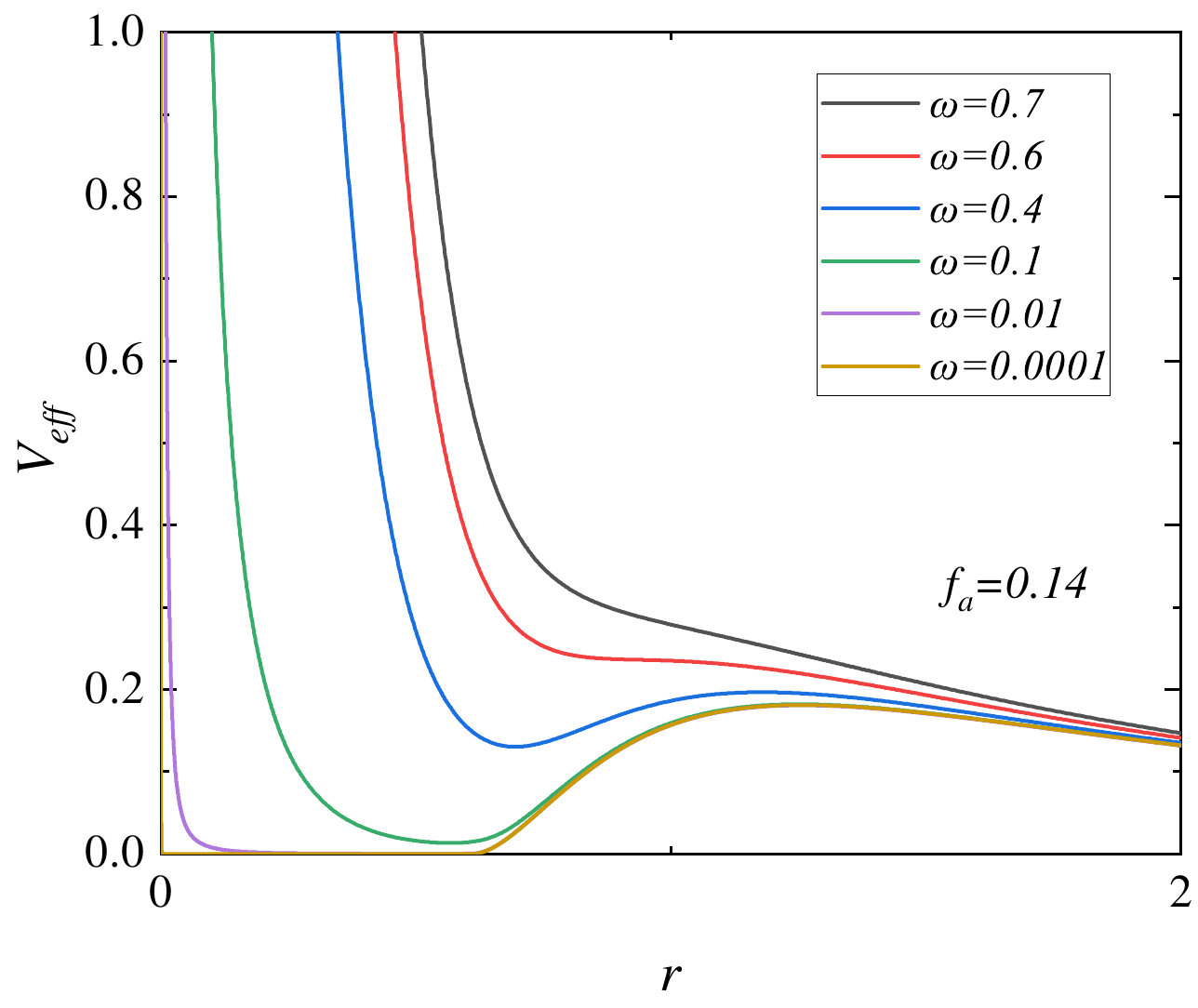}
  \includegraphics[width=0.45\textwidth]{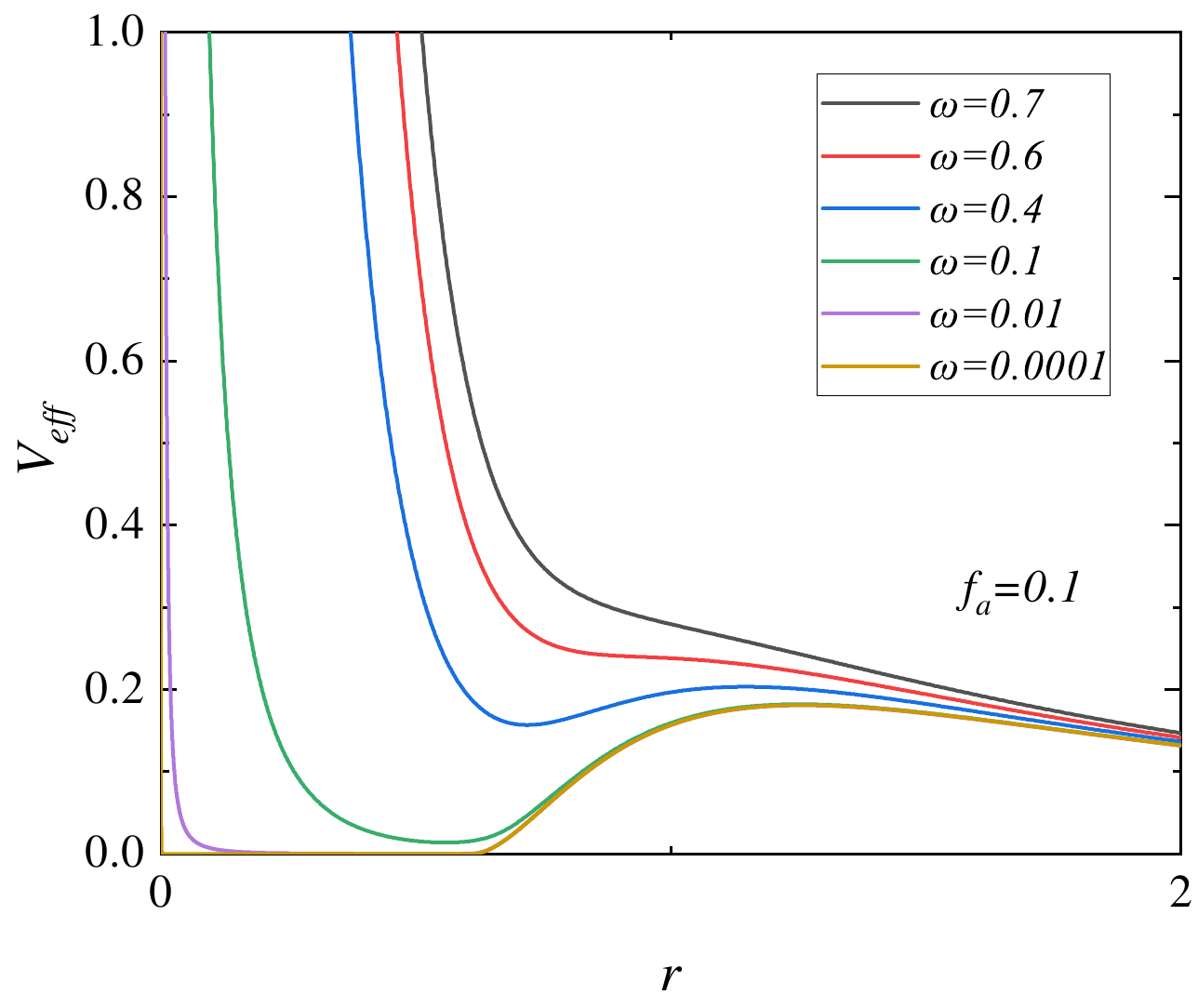}
  \includegraphics[width=0.45\textwidth]{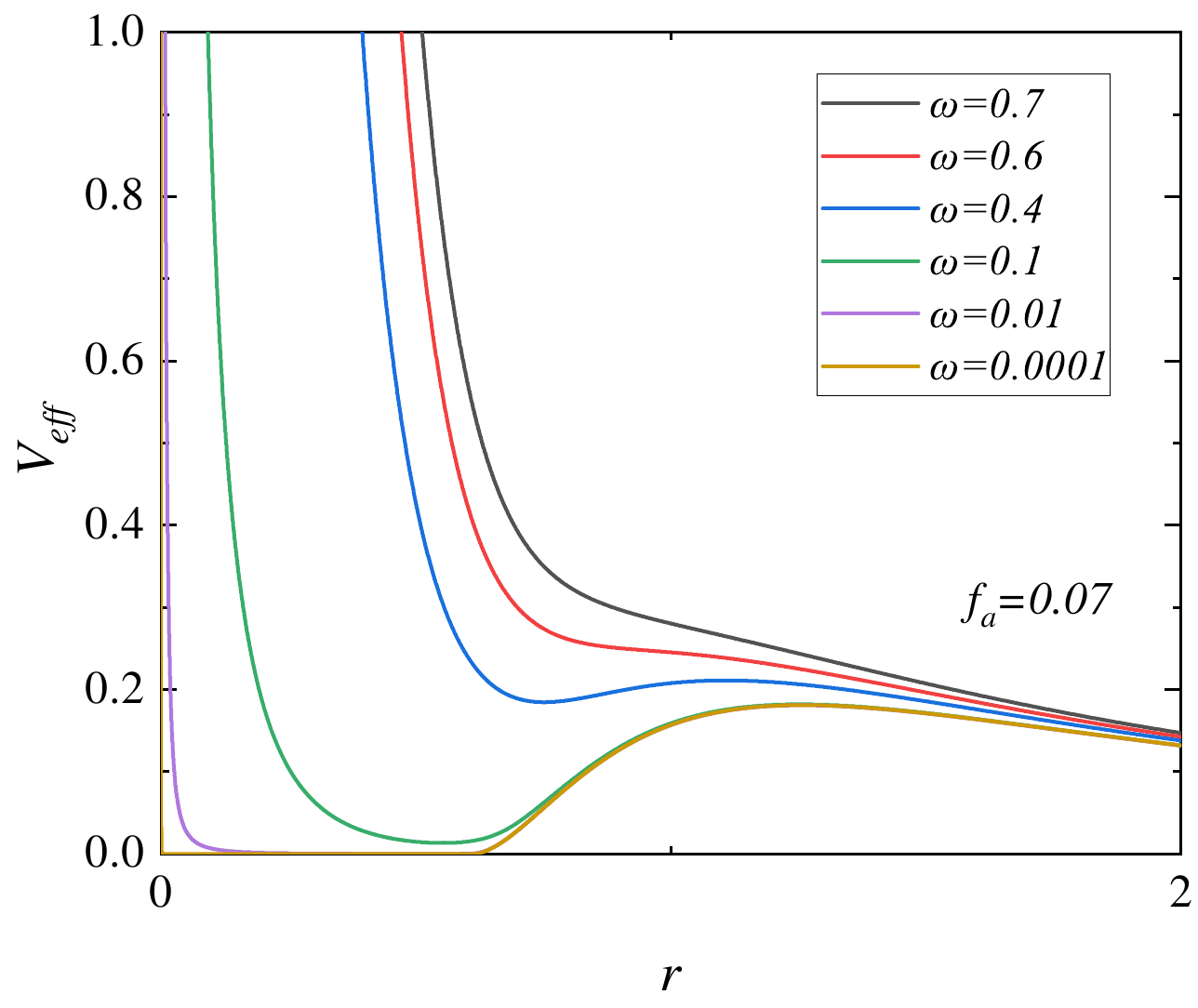}
   \includegraphics[width=0.45\textwidth]{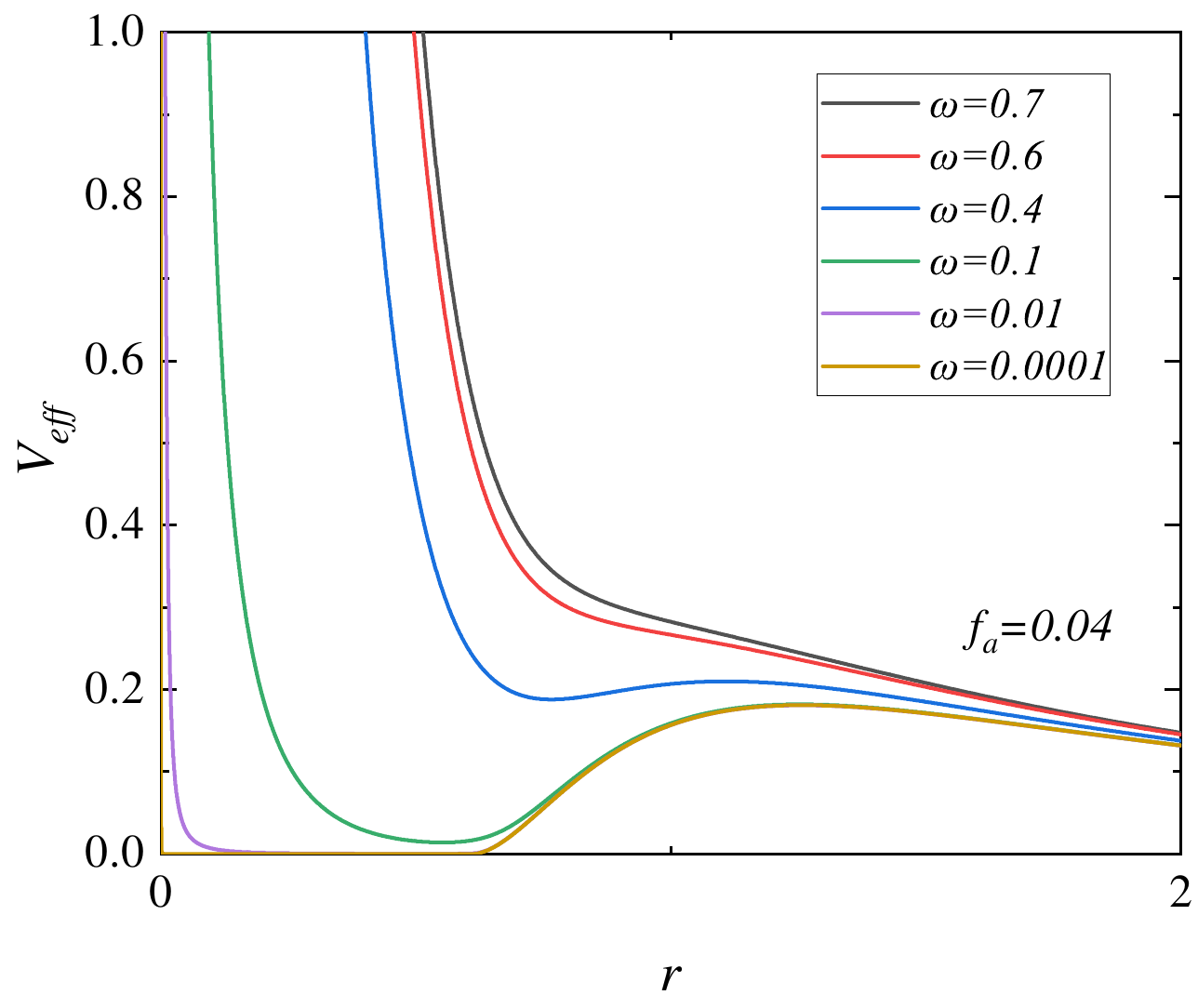}
  \includegraphics[width=0.45\textwidth]{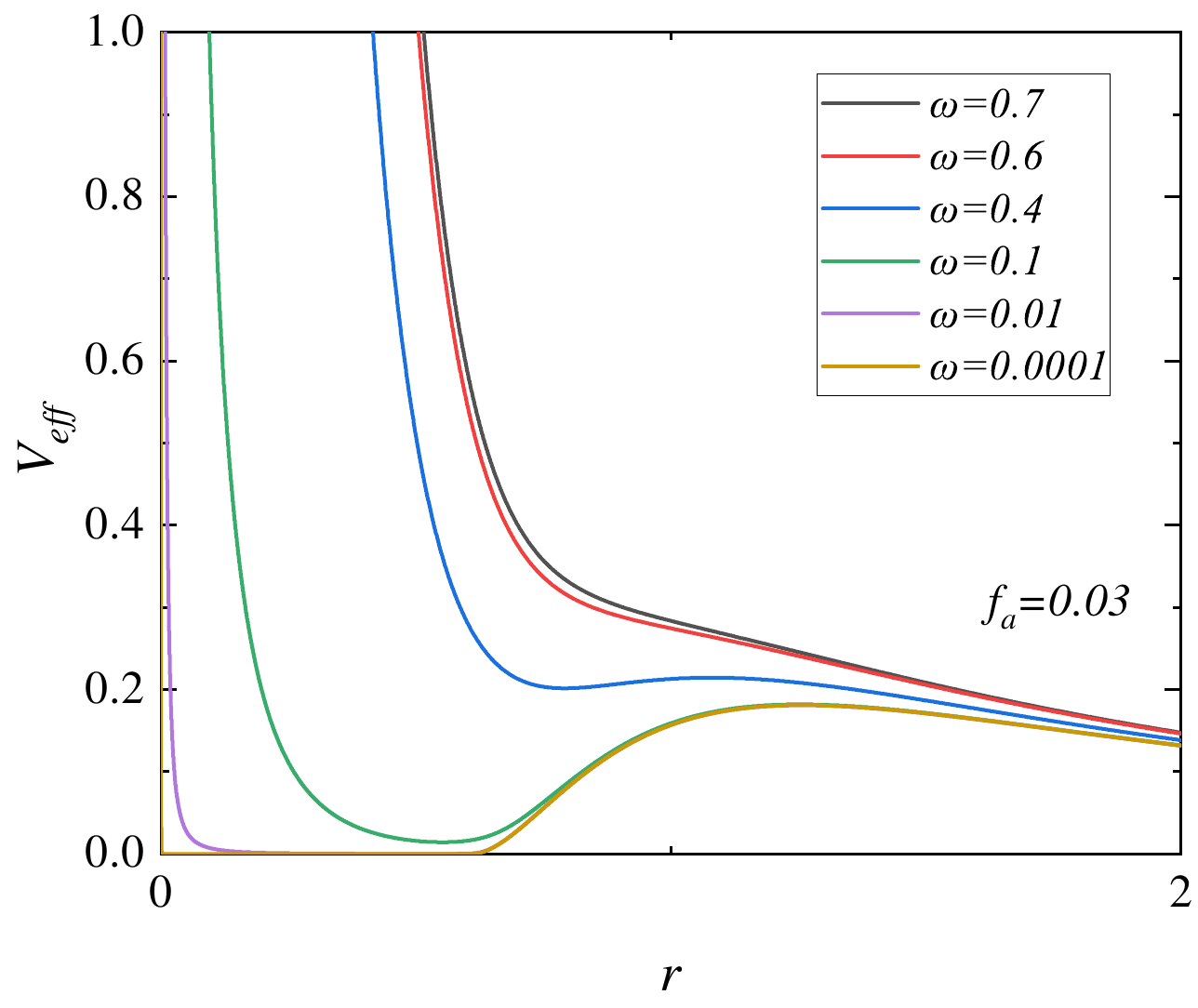}
  \caption{The distribution of the effective potential $V_{eff}$ along the radial coordinate $r$ for magnetic charge $q = 0.55$ with different decay constants $f_a$ at frequencies $\omega = \{0.7, 0.6, 0.4, 0.1, 0.01, 0.0001\}$.}
  \label{veff}
\end{figure}

In Fig.~\ref{veff}, we present the distribution of the effective potential along the radial coordinates for a magnetic charge $q=0.55$. Different frequencies are represented by solid lines of various colors. It is evident that the behavior of the effective potential varies with frequency.
When the frequency is high, there is no extreme point of the effective potential. As the frequency decreases, the effective potential exhibits one extremum point. With further decrease in frequency, the effective potential displays two extremum points. At $\omega=0.0001$, the effective potential exhibits two extremum points, indicating the presence of two light rings in the Hayward axion frozen star.

\begin{figure}[!htbp]
  \centering
  \includegraphics[width=0.49\textwidth]{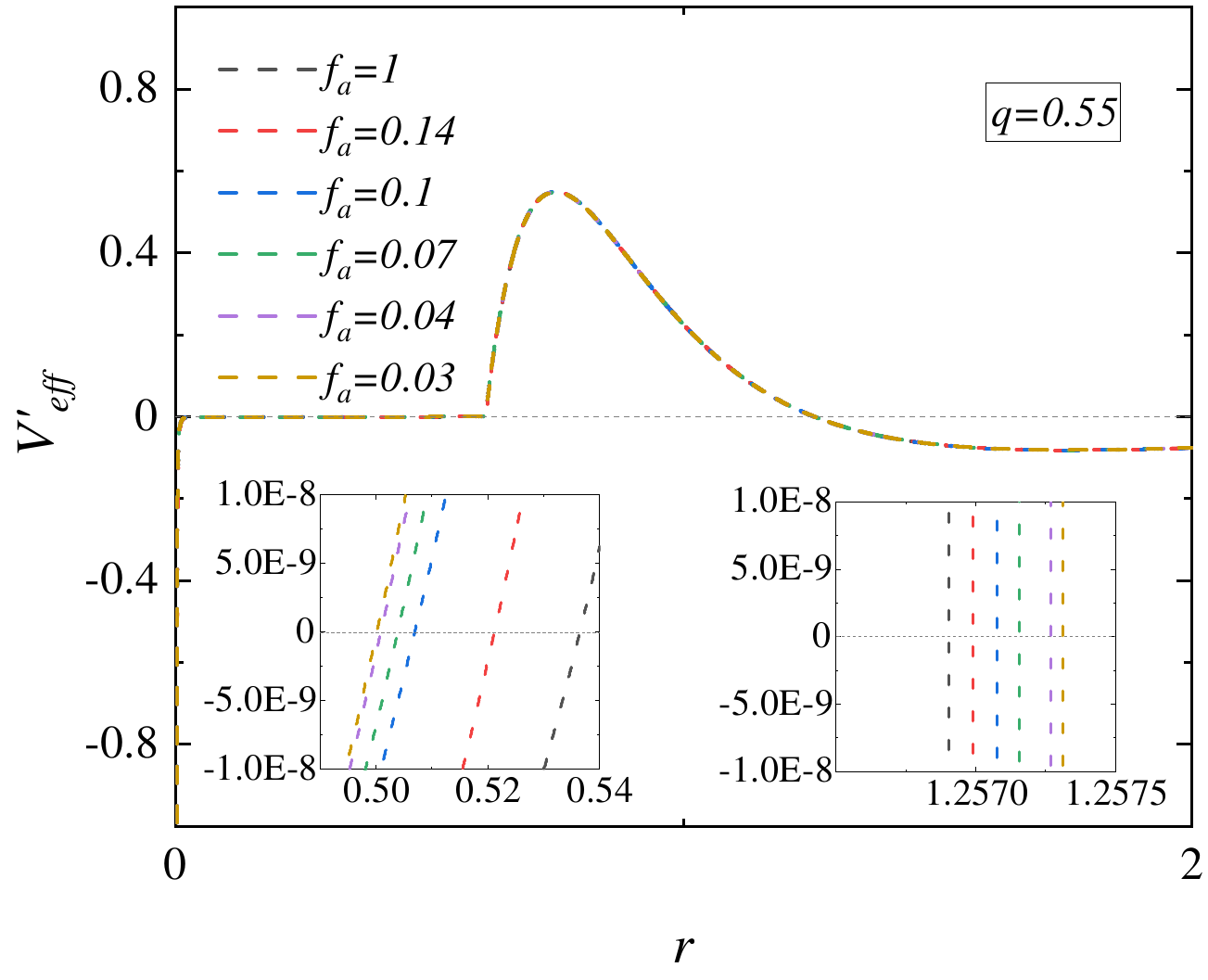}
  \includegraphics[width=0.49\textwidth]{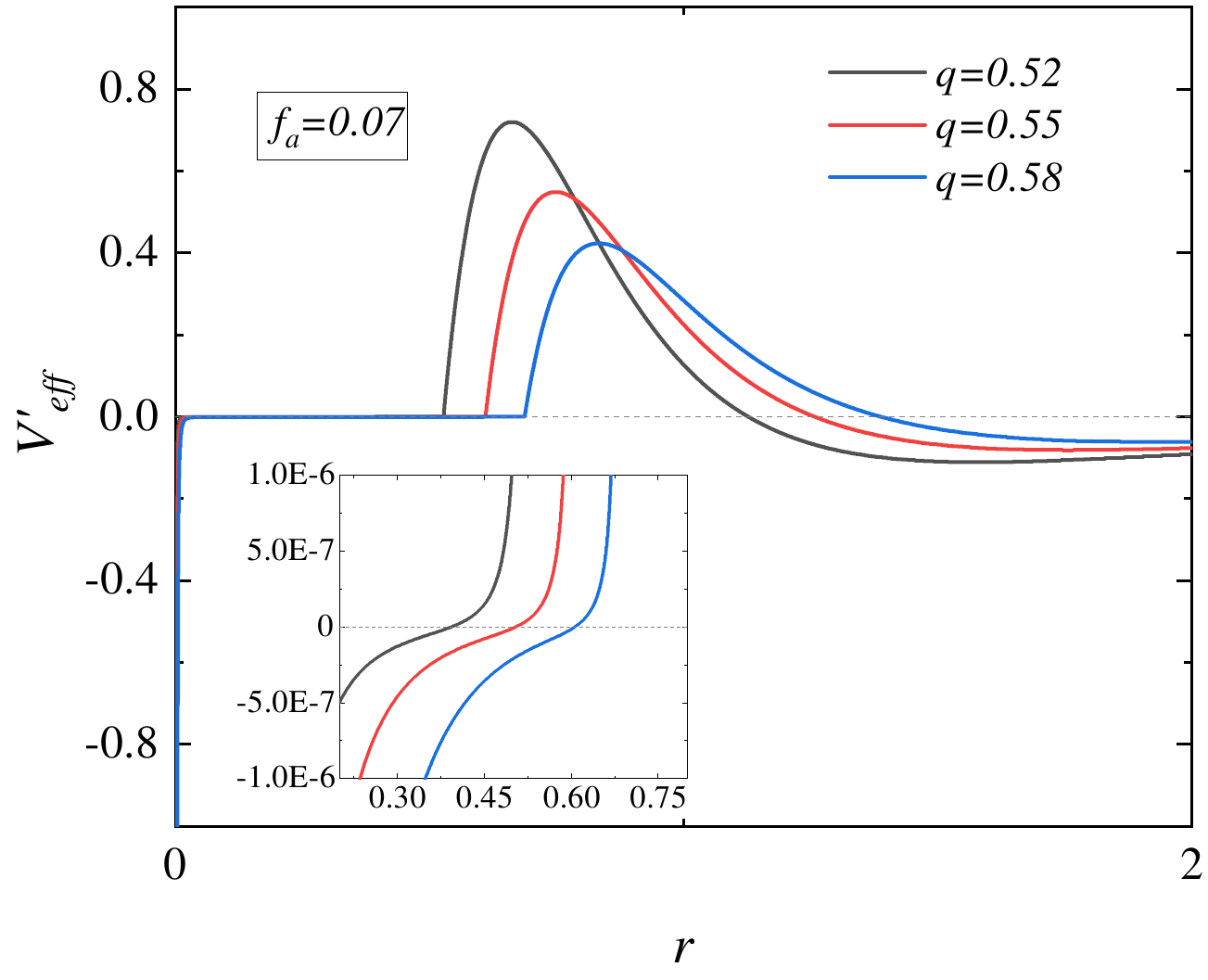}
  \caption{The distribution of the derivative of the effective potential $V'_{eff}$ along the radial coordinate $r$ at frequency $\omega = 0.0001$. The inset shows the details near $V'_{eff} = 0$. In the left panel, the magnetic charge is fixed at $q = 0.55$, and different decay constants $f_a = \{1, 0.14, 0.1, 0.07, 0.04, 0.03\}$ are selected. In the right panel, the decay constant is fixed at $f_a = 0.07$, and different magnetic charges $q = \{0.52, 0.55, 0.58\}$ are selected.}
  \label{veffd}
\end{figure}

To investigate the location of the light rings around the Hayward axion frozen stars, we present the distribution of the first derivative of the effective potential 
$V^{'}_{eff}$ 
  with respect to the radial coordinate 
$r$ at 
$\omega=0.0001$ in Fig.~\ref{veffd}. The inset highlights the details near $V^{'}_{eff}=0$, which indicate the positions of the light rings.
In the left panel, with the magnetic charge fixed at $q = 0.55$, we depict the distribution of the first derivative of the effective potential $V'_{eff}$ with respect to the radial coordinate $r$ for decay constants $f_a = \{1, 0.14, 0.1, 0.07, 0.04, 0.03\}$. 
The results indicate that variations in the decay constant $f_a$ have little impact on the position of the light rings.
The inner light ring shifts slightly inward as the decay constant $f_a$ decreases, while the position of the outer light ring remains nearly unchanged for different decay constants $f_a$.
In the right panel, with the decay constant fixed at $f_a = 0.07$, we illustrate the distribution of the first derivative of the effective potential $V'_{eff}$ with respect to the radial coordinate $r$ for magnetic charges $q = \{0.52, 0.55, 0.58\}$. 
It is observed that as the magnetic charge $q$ increases, both the inner and outer light rings of the Hayward axion frozen star move outward. 
Furthermore, for Hayward axion frozen stars, the inner light ring inside the critical horizon is stable, whereas the outer light ring outside the critical horizon is unstable. The first derivative of the effective potential $V'_{eff}$ approaches zero inside the critical horizon of the Hayward axion frozen star, implying that photons require a significant amount of time to reach the inner light ring.

     \section{Conclusion} \label{sec5}
In this paper, we construct a system with minimal coupling Einstein gravity, the nonlinear electromagnetic field, and the axion field, which is termed the Hayward axion star. Compared to Hayward black holes, Hayward axion stars successfully prevent the formation of the event horizon due to the presence of the complex scalar field.
Solutions for Hayward axion stars are classified according to the magnetic charge they possess. In scenarios where the magnetic charge is absent, the solutions correspond to conventional axion stars. When the magnetic charge remains below a critical threshold, the solutions do not exhibit zero frequency. Conversely, when the magnetic charge exceeds the critical value, a distinct category of solutions emerges, characterized by frequencies approaching zero, which are referred to as frozen stars.
In these frozen star solutions, the field function and energy density are primarily distributed within the critical horizon and drop sharply beyond it. Inside the critical horizon, the metric component $g_{tt}$ approaches zero, suggesting a near cessation of time. Furthermore, to distant observers, these solutions bear resemblance to extreme black holes.

However,  the axion field exhibits self-interaction, unlike a free scalar field, leading to distinct behaviors between Hayward axion frozen stars and Hayward boson frozen stars~\cite{Yue:2023sep}. 
Firstly, the magnetic charge necessary for the formation of a frozen star is inversely related to the decay constant of the axion star. As the decay constant decreases, the critical magnetic charge increases, suggesting that stronger self-interaction requires a higher magnetic charge for frozen star formation. Secondly, the mass of the Hayward axion frozen star is determined only by the magnetic charge and remains uninfluenced by the decay constant, indicating that the self-interaction does not influence the mass of the frozen star.

 Additionally, an analysis of the effective potential of Hayward axion stars reveals that at high frequencies, no light rings are present around the star. As the frequency decreases, a degenerate light ring forms, and with further reduction, two light rings emerge. In the case of Hayward axion frozen stars, two light rings are observed. The stable light ring appears inside the critical horizon, while the unstable light ring is located outside the critical horizon. Furthermore, it is noted that the decay constant 
$f_a$ has a relatively small impact on the positions of the light rings, whereas an increase in the magnetic charge 
$q$ causes the light rings to move outward. This outward shift is likely attributable to the alteration in the gravitational field induced by the increased magnetic charge, which subsequently affects the trajectory of light and the positions of stable orbits.

Hayward axion frozen stars present a promising approach to addressing the black hole singularity problem and the information loss paradox. These stars avoid the formation of an event horizon, as seen in traditional black holes, thereby theoretically allowing for the preservation and transmission of information.  In future work, it would be highly meaningful to investigate whether Hayward axion frozen stars exhibit Hawking radiation. Additionally, the stability of Hayward axion stars is an issue that deserves to be explored in depth. Understanding the dynamic stability of these stars will be essential in assessing their feasibility and physical plausibility as alternative black hole models. Such studies will not only enhance our comprehension of Hayward axion stars but also potentially offer new insights into the unification of general relativity and quantum mechanics.

	\section*{ACKNOWLEDGEMENTS}
	This work is supported by National Key Research and Development Program of China (Grant No. 2020YFC2201503) and the National Natural Science Foundation of China (Grants No.~12275110 and No.~12047501). Parts of computations were performed on the Shared Memory system at Institute of Computational Physics and Complex Systems in Lanzhou University.

\end{document}